\documentclass[reprint, superscriptaddress, amsmath,amssymb, aps, pra, groupedaddress, longbibliography]{revtex4-2}

\usepackage{color}

\usepackage{natbib}
\usepackage{bbold}
\usepackage{comment}
\usepackage{booktabs}

\newcommand{\me}{\operatorname{e}}
\newcommand{\dager}{^{\dagger}}
\renewcommand{\a}{\hat{a}}
\renewcommand{\b}{\hat{b}}
\newcommand{\cc}{\hat{c}_{\bm{k},\bm{e}}}
\newcommand{\cl}{\hat{c}_l}
\newcommand{\RR}{\bm{\hat{R}}}
\newcommand{\rr}{\hat{\bm{r}}}

\newcommand{\PP}{\bm{\hat{P}}}
\newcommand{\x}{\hat{x}}

\newcommand{\z}{\hat{z}}
\newcommand{\p}{\hat{p}}

\newcommand{\Q}{\hat{Q}}
\renewcommand{\P}{\hat{P}}
\renewcommand{\Q}{\hat{Q}}
\renewcommand{\H}{\hat{H}}

\newcommand{\U}{\hat{U}}

\newcommand{\ra}{\rightarrow}
\newcommand{\overbar}[1]{\mkern 1.5mu\overline{\mkern-1.5mu#1\mkern-1.5mu}\mkern 1.5mu}

\newcommand{\half}{\frac{1}{2}}
\renewcommand{\r}{\bm{r}}
\newcommand{\eye}{\mathbb{1}}
\newcommand{\tw}{_{ \mathrm{t}}}
\newcommand{\twi}{_{ \mathrm{t},i}}
\newcommand{\twj}{_{ \mathrm{t},j}}
\newcommand{\thj}{_{\mathrm{th}, j}}

\newcommand{\cav}{_{\mathrm{c}}}

\newcommand{\envj}{_{\mathrm{env}, j}}

\newcommand{\free}{_{\mathrm{f}}}

\usepackage{epsfig} 
\usepackage[dvipsnames]{xcolor}

\usepackage{lipsum}

\usepackage{graphicx}
\usepackage{dcolumn}
\usepackage{bm}
\usepackage[parse-numbers=false]{siunitx}
\usepackage{upgreek}

\colorlet{shadecolor}{gray!40}

\begin{document}

\preprint{APS/123-QED}

\title{Coherent Scattering-mediated correlations between levitated nanospheres} 

\author{I. Brand\~{a}o}
\email{igorbrandao@aluno.puc-rio.br}
\affiliation{Departamento de F\'{i}sica, Pontif\'{i}cia Universidade Cat\'{o}lica do Rio de Janeiro,  22451-900 Rio de Janeiro, RJ, Brazil}
\author{D. Tandeitnik}
\email{tandeitnik@aluno.puc-rio.br}
\affiliation{Departamento de F\'{i}sica, Pontif\'{i}cia Universidade Cat\'{o}lica do Rio de Janeiro,  22451-900 Rio de Janeiro, RJ, Brazil}
\author{T. Guerreiro}
\email{barbosa@puc-rio.br}
\affiliation{Departamento de F\'{i}sica, Pontif\'{i}cia Universidade Cat\'{o}lica do Rio de Janeiro,  22451-900 Rio de Janeiro, RJ, Brazil}

\date{\today}

\begin{abstract}
We explore entanglement generation between multiple optically levitated nanospheres interacting with a common optical cavity via the Coherent Scattering optomechanical interaction. We derive the many-particle Hamiltonian governing the unitary evolution of the system and show that it gives rise to quantum correlations among the various partitions of the setup, following a non-Markovian dynamics of entanglement birth, death and revivals. 
We also consider the effects of coupling the system to external environments and show that under reasonable experimental conditions entanglement between the mechanical modes can survive even at room temperature. Its dependence upon the number of nanoparticles, their initial temperature and coupling strength is studied. A numerical toolbox to simulate the closed and open dynamics of Gaussian optomechanical states and their informational measures is developed.
\end{abstract} 

\maketitle


\section{Introduction}\label{sec:introduction}

Testing quantum mechanics in novel regimes, such as observing quantum effects in systems with many constituents or a large number of degrees-of-freedom, is one of the cornerstones of fundamental science and a promising achievement towards new technologies. A number of experiments have contributed along that direction by studying the quantum mechanics of nano- and microscale objects. For instance, entanglement of hundreds of ions has been observed and controlled \cite{Bohnet2016}, interferometric systems have achieved micron-spaced superposition of atomic wavefunctions \cite{Xu2019}, coherence in Bose-Einstein condensates has been observed \cite{Berrada2013} and ground state cooling and coherent control of micron-sized cantilevers and their coupling to superconducting quantum electronics proposed \cite{Huo2008,Etaki2008} and demonstrated \cite{OConnell2010}. 

Optically levitated nanoparticles allow exceptional control over translational \cite{Chan2011, Delic2020a, Magrini2020} and rotational \cite{Kuhn2017, Schafer2020, Stickler2021} degrees-of-freedom and achieve excellent environmental isolation \cite{Millen2020a, Millen2020}, thus providing a promising setup for pushing the boundaries of quantum theory towards unexplored regimes.
Proposals for generating spatial superposition of levitated nanoparticles have been put forward \cite{RomeroIsart2010, RomeroIsart-superposition-2011, Weiss2020}, as well as for testing collapse models \cite{Carlesso2020} and witnessing nonclassicality through recurrence of optical squeezing \cite{Ma2020} and optical entanglement \cite{Brandao2020}. Moreover, levitated systems can give rise to steady-state entanglement \cite{Paternostro2007,Vitali2007, Krisnanda2020, Gut2020, Chauhan2020} and help in the search for new physics \cite{Aggarwal2020searching}.
On the experimental front, the possibility of detecting nonclassical correlations in levitated particles has been demonstrated \cite{Rakhubovsky2020}. Effective 3D cooling \cite{Windey2019, Delic2019}, ground state cooling \cite{Delic2020a, Magrini2020} and strong light-matter coupling have been realized \cite{DeLosRios2020}. All of these are essential requirements towards entering and controlling the mesoscopic quantum regime.

Ground state cooling of levitated nanoparticles along a single axis  was first enabled through the so-called coherent scattering interaction \cite{Vuletic2001, Delic2020a}, and it has been theoretically shown that simultaneous $2$D ground state cooling is possible with the same technique \cite{Toros2020}. In this cooling scheme, motion of the particle coherently scatters photons from the trapping beam into an optical cavity tuned to enhance scattering of photons that carry away energy from the trapped object \cite{Gonzalez-Ballestero2019, Delic2020a}. 
Following the recent interest on entanglement dynamics in optomechanical systems \cite{Cheng2016, Chen2017,Li2019, Brandao2020, Faroughi2021}, coherent scattering has also been considered as a platform for generating mechanical entanglement \cite{Chauhan2020, Hornberger2020}. 
The present work builds along that direction and investigates how the coherent scattering interaction between a single cavity mode and an arbitrary number of levitated nanoparticles can give rise to quantum correlations among the various partitions of the system even at room temperature.

We begin by deriving the many-particle coherent scattering Hamiltonian in close analogy to \cite{Gonzalez-Ballestero2019, Romero-Isart2011}, where an arbitrary number of nanoparticles share a common optical cavity. By appropriate positioning of the particles with respect to the cavity nodes one can minimize the dispersive optomechanical interaction and favor the coherent scattering terms.
The unitary dynamics generated by the Hamiltonian is responsible for creating entanglement in the system. We discuss the closed system evolution through numerical simulation of the so-called Lyapunov equation in the absence of photon loss and decoherence. In the unitary regime, we show the occurrence of periodic entanglement birth, death and revivals evidencing the non-Markovian nature of the evolution. 
Real-life quantum systems, however, are open. For this reason, we model the environmental interactions through a set of quantum Langevin equations and the associated Lyapunov equation; both closed and open dynamics are simulated with a custom numerical toolbox \cite{numerical_toolbox}. We proceed to study entanglement generation in the presence of collisional decoherence and photon recoil heating. 
We demonstrate the persistence of entanglement oscillations and non-Markovianity even in the presence of noise and contact to bosonic and heat baths, and discuss the prospects of experimentally verifying these quantum correlations with current technology.

\section{Hamiltonian}
\label{sec:Hamiltonian}

\begin{figure}[t]  
    \centering
    \includegraphics[width=8.6cm]{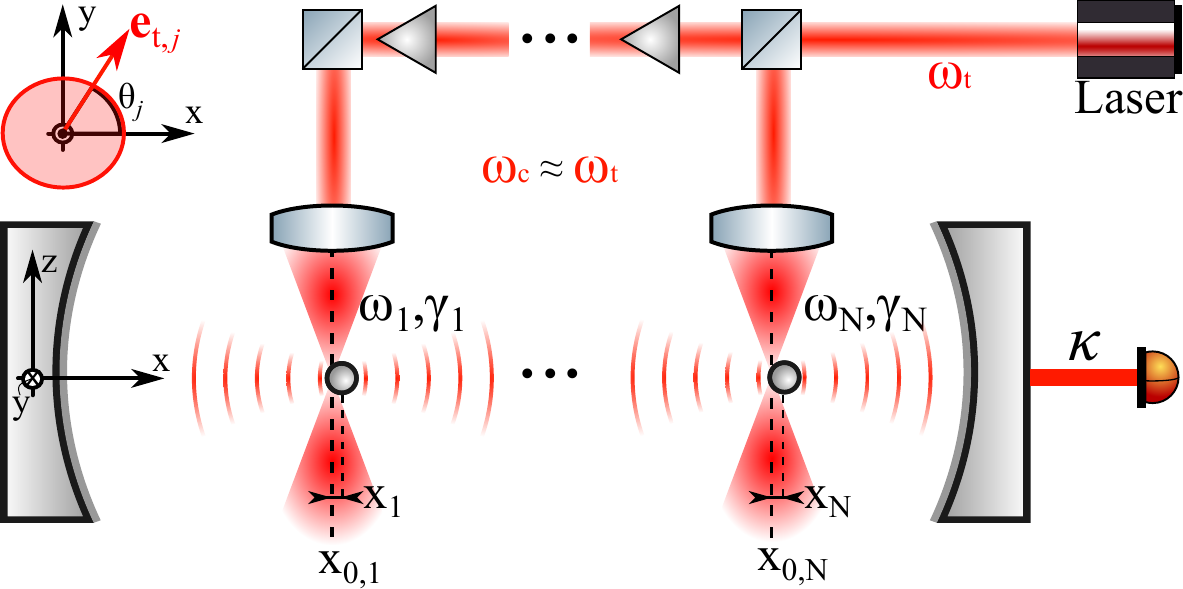}
    \caption{Schematics of $N$ optically levitated nanoparticles, and its corresponding tweezers inside a common optical cavity. Every tweezer is considered to have the same frequency, which is tuned close to the cavity resonance frequency, such that the scattered photons can survive inside the cavity. It is assumed that the tweezers are sufficiently spaced apart such that beam overlap can be neglected. 
    Information about the system can be retrieved through the leaking field from the rightmost end mirror of the cavity.}
    \label{fig:CS_complete}
\end{figure}

The system we are interested in is comprised of $N$ optically trapped dielectric nanoparticles (NP) of mass $m_j$, each with radius $R_j$ on the order of magnitude of $\SI{100}{\nano\meter}$, refractive index $n_{\mathrm{R},j}$, homogeneous and isotropic permitivitty $\epsilon_j \approx n_{\mathrm{R}, j}^2$, and polarizability $\overbar{\alpha}_j \equiv 3\epsilon_0\frac{\epsilon_j - 1}{\epsilon_j + 2}$. Each NP is optically trapped by an independent optical tweezer (OT) and placed on the axis of a Fabry-Pérot cavity of length $L$ and resonance frequency $\omega\cav$, as depicted in Figure \ref{fig:CS_complete}. The tweezers are assumed to be sufficiently apart such that any overlap and cross-talk between the traps can be neglected.
All OTs propagate perpendicularly to the cavity axis, have the same frequency $\omega\tw = 2\pi c /\lambda\tw$ and their polarization vectors $\bm{e}\twj$ can be decomposed as $\bm{e}\twj = \cos(\theta_j)\bm{e}_{x} + \sin(\theta_j)\bm{e}_{y}$ along the cavity axis.

Following \cite{Gonzalez-Ballestero2019}, the total Hamiltonian governing the system dynamics can be written as
\begin{equation}
    \label{eq:total_hamiltonian} \H = \H_{\mathrm{NP}} + \H_{\mathrm{field}} + \H_{\mathrm{int}} \, ,
\end{equation}
\noindent where the first term is the energy of the free NPs, the second term is the total energy stored in the electromagnetic (EM) field and the third term represents interactions between the NPs and the EM field. To place this Hamiltonian in a suitable form, we consider the different contributions from the electric field present in the system and work on approximating each of these terms individually.


The total electric field is considered to be approximately given by a sum of contributions from the fields of each OT, $\bm{E}\twj$, and  the intracavity field $\bm{\hat{E}}\cav$,
\begin{align}\label{eq:total_em_field}
    \bm{\hat{E}}(\bm{r}) \simeq \bm{\hat{E}}\cav(\bm{r}) + \sum_{j}^{N}\bm{E}\twj(\bm{r}) \, . 
\end{align}
The OTs are considered to be in strong coherent states \cite{Gonzalez-Ballestero2019} and thus well described by a classical field. The mean value of the $j$-th OT's electric field operator in an appropriate rotating frame is given by
\begin{align}
    &\bm{E}\twj(\bm{r}, t)  \label{eq:tweezer_electric_field} =  \\& \half \epsilon\tw\frac{w_{0,j}}{w_j(z)}  \me^{-\frac{(x-x_{0,j})^2 + y^2}{w_{0,j}^2}}\me^{ik\tw z}\me^{i\phi_{G,j}(z)}\me^{i\omega\tw t}\bm{e}\tw+ \mathrm{c.c.} ,  \nonumber
\end{align}
\noindent where $P\twj$ is the power, $k\tw = 2\pi/\lambda\tw$ the wave-number, $\bm{e}\twj$ the polarization vector, $w_{0,j}$ the waist and $ \epsilon\twj ,  w_j(z), \phi_G(z), z_{\mathrm{R},j} $ are the field amplitude, beam width, Gouy phase and Rayleigh range, respectively. These quantities are given by
\begin{eqnarray}
    \epsilon\twj &=& \sqrt{\frac{4P\twj}{w_{0,j}^2 \pi \epsilon_0 c}} \, ,\\ \nonumber
    w_j(z) &=& w_{0,j}\sqrt{1 + z^2/z_{\mathrm{R},j}^2} \, ,\\ \nonumber
    \phi_G(z) &=& - \arctan(z/z_{\mathrm{R},j}) \, , \\ \nonumber
    z_{\mathrm{R},j} &=& k\tw w_{0,j}^2/2 \, .
\end{eqnarray}
The intracavity electric field is a standing wave described quantum mechanically by the operator 
\begin{equation}
    \hspace*{-3mm}\bm{\hat{E}}\cav(\bm{r}) =  \epsilon\cav \left( \a\dager + \a \right) \cos(k\cav x - \phi)\bm{e}_{y}, \label{eq:cavity_electric_field}
\end{equation}
\noindent where $k\cav = 2\pi/\lambda\cav$ is the wave vector, $\epsilon\cav = \sqrt{\frac{\hbar\omega\cav}{2\epsilon_0 V\cav}}$ the single photon electric field for a cavity of mode volume $V\cav$, $\a$ the time-dependent annihilation operator, $\bm{e}_{y}$ the cavity field polarization and $\phi=\pi/2$ due to our choice of coordinates in Figure \ref{fig:CS_complete}. We note that the external free EM field also plays a role in the dynamics of the NPs. However, as shown in the Appendix \ref{appendix:free_field}, the effect of interaction with this field is negligible if the NPs are properly positioned within the cavity and sufficiently cooled down. We will therefore drop any term involving the free EM field in what follows.


In the long wavelength approximation, given by $R_j \ll \lambda\cav, \lambda\twj$, the interaction Hamiltonian can be expressed as~\cite{Gonzalez-Ballestero2019, Romero-Isart2011, Jackson1998},
\begin{align}
    \H_{\rm int} &= -\half\int \bm{P}(\bm{r})\bm{E}(\bm{r}) d^3\bm{r} 
    \simeq -\half\sum_{j=1}^N\alpha_j|\bm{\hat{E}}(\rr_j)|^2 \label{eq:Int_Hamiltonian} \\ &\simeq -\half\sum_{j=1}^N\alpha_j\bigg\vert\bm{\hat{E}}\cav(\rr_j)  + \bm{E}\twj(\rr_j)\bigg\vert^2 \nonumber \, , 
\end{align}
\noindent where $\bm{P}_j(\bm{r}) = \overbar{\alpha}_j\bm{E}(\bm{r})$ is the $j$-th NP polarization vector and $\alpha = \overbar{\alpha}V = 4\pi\epsilon_0 R^3\frac{\epsilon_j - 1}{\epsilon_j + 2}$. Here $\rr_j = \bm{R}_{0,j} + \RR_j$ denotes the center-of-mass (COM) position operator of the $j$-th particle, with $\bm{R}_{0,j} = (x_{0,j} , 0, 0)$ being the mean position of the $j$-th OT along the cavity axis and $\RR_j = (\hat{X}_j, \hat{Y}_j, \hat{Z}_j)$ the fluctuations of the particle around $\bm{R}_{0,j}$. The interaction Hamiltonian is simplified by assuming that the \textit{overlap} term proportional to $\bm{E}\twi(\rr_j)$ is negligible for $i\neq j$, i.e., $\bm{E}\twi(\rr_j) \approx \delta_{ij}\bm{E}\twi(\rr_j)$. 

Terms proportional to $\vert\bm{E}\twj(\rr_j)\vert^2$ give rise to  a $3D$ harmonic potential on the NPs, effectively levitating them with trapping frequencies $\omega_{\alpha,j}$, $\alpha=x,y,z$, given by
\begin{equation}
    \begin{bmatrix} \omega_{x,j} \\ \omega_{y,j} \\ \omega_{z,j} \end{bmatrix} = \begin{bmatrix} \sqrt{\frac{4\alpha_j P\twj}{m_j w_{0,j}^4 \pi \epsilon_0 c}} \\ \sqrt{\frac{4\alpha_j P\twj}{m_j w_{0,j}^4 \pi \epsilon_0 c}} \\ \sqrt{\frac{2\alpha_j P\twj}{m_j  w_{0,j}^2  z_{\mathrm{R},j}^2 \pi \epsilon_0 c}} \end{bmatrix}.\label{eq:trapping_frequencies}
\end{equation}
\noindent Moreover, terms proportional to $\vert\bm{E}\cav(\rr_j)\vert^2$ result in three contributions: a shift in the natural frequency of the cavity, 
a radiation pressure coupling and a driving term on the $j$-th NP. 

Finally, the last term proportional to $\text{Re}\big\{\bm{E}\twj(\rr_j)\bm{E}\cav(\rr_j)\big\}$ generates the coherent scattering (CS) interaction \cite{Gonzalez-Ballestero2019}, effectively 2D coupling the NPs with the cavity field, and a drive term in the cavity field. The nature of this interaction can be intuitively understood as due to the OT's photons being coherently scattered by the trapped NPs into the cavity mode, thus populating it \cite{Vuletic2001, Gonzalez-Ballestero2019, Delic2019}.

Gathering all the terms, disregarding constant energy shifts and moving into the appropriate rotating frame at the frequency of the OTs (see Appendix \ref{sec:appendix_Hamiltonian_details} for more details), we find the full Hamiltonian for the CS system with $ N$ particles: 
\begin{widetext}
\begin{align}
    &\H = \hbar \Delta \a\dager\a + \sum_{j=1}^{N}\Bigg\{ \frac{\bm{\P}_{j}^2}{2m_j} + \RR_j^T \Omega_j ^2\RR_j - \hbar G_j \sin(\theta_j)\cos\big(k\cav x_{0,j} - \phi\big) \big( \a\dager + \a \big) - \hbar \frac{g_{0,j}}{2}\sin\big[2(k\cav x_{0,j} - \phi)\big]\hat{X}_j \nonumber \\ 
    &- \hbar g_{0,j}\sin\big[2(k\cav x_{0,j}-\phi)\big]\a\dager\a\,\hat{X}_j 
    + \hbar G_{x,j}\sin\big(k\cav x_{0,j} - \phi\big)\sin(\theta_j)\big( \a\dager + \a \big) \, \hat{X}_j + i\hbar G_{z,j}\cos\big(k\cav x_{0,j} - \phi\big)  \sin(\theta_j)\big( \a\dager - \a \big) \hat{Z}_j \Bigg\} \label{eq:Full_Hamiltonian}  \, , 
\end{align}
\end{widetext}
\noindent where $\Omega_j \equiv \text{diag}(\omega_{x,j}, \omega_{y,j}, \omega_{z,j})$ describe the trap frequencies for the j-th NP, $\Delta \equiv \omega\cav -\omega\tw - \sum_{j=1}^{N}\delta_j \cos^2(k\cav x_{0,j} - \phi)$ is the shifted cavity frequency, $g_{0,j} = k\cav\delta_j$ are the dispersive couplings with $\delta_j = \frac{\alpha_j\omega\cav}{2\epsilon_0 V\cav}$, $G_{x,j} = k\cav G_j$ and $G_{z,j} = \frac{k\tw z_{\mathrm{R},j} - 1}{z_{\mathrm{R},j}}G_j$ are the bare CS couplings in the $x$ and $z$ directions for the $j$-th particle with $G_j = \alpha_j\epsilon\twj \epsilon\cav/(2\hbar)$, and $\PP_j = (\hat{P}_{x,j}, \hat{P}_{y,j}, \hat{P}_{z,j})$ is the momentum operator for the $j$-th particle. 



In order to favour the CS couplings over the dispersive ones, we place the NPs' mean position at the cavity nodes by acting on their respective OT. Consequently the total Hamiltonian simplifies to

\begin{align} 
    \hspace*{-1em} \H/\hbar \approx  \Delta \a\dager\a + \sum_{j=1}^{N}  \omega_j \b_j\dager\b_j 
    + \sum_{j=1}^{N}  g_j\,(\a\dager+\a)(\b_j\dager+\b_j) \, , \label{eq:CS_Hamiltonian_final}
\end{align}

\noindent  where $\b_j (\b_j)$ is the annihilation (creation) operator for the $j$-th NP. Note that the shifted cavity frequency simplifies to $\Delta = \omega\cav - \omega\tw$ and $g_j\equiv x_{\mathrm{ZPF}, j} G_{x,j}\sin(\theta_j)$ is the CS coupling, with $x_{\mathrm{ZPF}, j} = \sqrt{\hbar / ( 2m_j \omega_j)}$ the zero point fluctuation for the $j$-th NP.

Note that for the case of a single NP, the Hamiltonian is symmetric between optical and mechanical modes. We note that it describes also the situation within the linearized dispersive optomechanical approximation \cite{Review_Aspelmeyer}, as well as the linearized membrane-membrane coupling \cite{Chen2020}. 

\section{Equations of Motion} \label{sec:open_dynamics}

The Hamiltonian \eqref{eq:CS_Hamiltonian_final} dictates closed unitary dynamics. 
We now proceed to discuss the open quantum dynamics of the system through a set of quantum Langevin equations \cite{Giovannetti2001, Vitali2007, Genes2008, Genes2009, Sommer2019}. We consider that one of the cavity mirrors is not perfect, resulting in a finite cavity linewidth $\kappa$ and allowing photon exchange between the cavity field and the external free field \cite{Gardiner1985}. Moreover, each NP is considered to be in contact with its own thermal bath, at temperature $T_{\mathrm{env}, j}$, resulting in a quantum Brownian motion for the NP \cite{Vitali2001}. We define the dimensionless position and momentum quadratures for each particle $\x_j = \b_j\dager + \b_j$, $\p_j = i(\b_j\dager - \b_j)$, and for the cavity field $\Q = \a\dager + \a$, $\P = i(\a\dager - \a)$, such that the quantum Langevin equations read
\begin{align}
    \dot{\Q}\,\; &= +\Delta\P - \frac{\kappa}{2}\Q + \sqrt{\kappa}\,\x_{\rm in} \, , \label{eq:Qtimeevo} \\
    \dot{\P}\,\; &= -\Delta\Q - \frac{\kappa}{2}\P + \sqrt{\kappa}\,\p_{\rm in} - \sum_{j=1}^N 2g_j\x_j \, , \label{eq:Ptimeevo} \\
    \dot{\x}_j &= +\omega_j\p_j \, , \label{eq:xtimeevo} \\
    \dot{\p}_j &= -\omega_j\x_j -\gamma_j\p_j + \hat{f}_j - 2g_j\Q \, , \label{eq:ptimeevo}
\end{align} 
\noindent where $\x_{\rm in}=\a_{\mathrm{in}}\dager + \a_{\mathrm{in}}$, $\p_{\rm in}=i(\a_{\mathrm{in}}\dager - \a_{\mathrm{in}})$ are the zero-averaged delta-correlated optical input noise terms satisfying $\langle \a_{\mathrm{in}}(t) \a_{\mathrm{in}}\dager(t') \rangle = \delta(t-t')$ \cite{Gardiner1985}; $\gamma_j$ is the damping rate for the $j$-th NP, which is under the influence of zero-averaged stochastic thermal noise $\hat{f}_j$ \cite{Vitali2001} with correlation functions given by
\begin{align*}
\langle &\hat{f}_j(t) \hat{f}_k(t')\rangle \\ &= \frac{2\gamma_j}{\omega_j} \int_{0}^{\overbar{\omega}_j}  \me^{-i\omega(t-t')} \omega\bigg[\coth\left(\frac{\hbar\omega}{2k_BT_{\mathrm{env}, j}}\right) + 1 \bigg] \frac{d\omega}{\pi} \, \delta_{j,k}
\end{align*}
\noindent Here, the reservoir cut-off frequency is $\overbar{\omega}_j$, and $k_B$ denotes the Boltzmann constant. From the form of these correlators, the stochastic noise automatically satisfies the fluctuation-dissipation theorem and in the high-temperature regime, $k_B T_{\mathrm{env}, j} \gg \hbar \omega_j$, becomes delta-correlated $\langle \hat{f}_j(t) \hat{f}_k(t') \rangle \approx  2\gamma_j(2\overbar{n}\thj + 1) \, \delta(t-t')\, \delta_{j,k}$ \cite{Genes2008, Genes2009, Vitali2001}, where $\overbar{n}\thj = \big[  1 - \exp[\hbar\omega_j / (k_B T_{\mathrm{env}, j})] \big]^{-1} \approx k_B T_{\mathrm{env}, j}/ (\hbar\omega_j)$ is the thermal occupation number for the $j$-th heat bath.

We are interested in the case where the $j$-th NP is initially in a thermal state at temperature $T_j$ with occupation number $\overbar{n}_{0,j}$ and the cavity field starts in the vacuum state.
The linear nature of the Langevin equations preserves the Gaussianity of the initial states, allowing the use of the Gaussian quantum information toolbox \cite{Lloyd2012}.
In particular, since Gaussian quantum states are completely characterized by their first and second moments, we can focus directly on the dynamics of the covariance matrix. See Appendix \ref{appendix:gaussian_toolbox} for a brief introduction to the structure of Gaussian states and definitions of useful quantities  used in remaining of this work such as the logarithmic negativity, von Neumann entropy and mutual information. 

Note that first moments can be easily calculated once we recast the Langevin equations in a more compact form, 
\begin{equation}
    \dot{\bm{\hat{X}}}(t) = A\bm{\hat{X}}(t) + \bm{\hat{N}}(t) \, ,
\end{equation}
\noindent where $\bm{\hat{X}} = (\Q, \P, \x_1, \p_1, ...)^T$ is the quadrature vector, $\bm{\hat{N}} = (\sqrt{\kappa}\x_{\rm in}, \sqrt{\kappa}\p_{\rm in}, 0, \hat{f}_1, ...)^T$ is the input noise vector and $A$ is the \textit{diffusion} matrix. 


Consider the formal expression for the quadratures 
\begin{equation}
    \bm{\hat{X}}(t) = \me^{A(t-t_0)}\bm{\hat{X}}(t_0) + \int_{t_0}^{t} \me^{A(t-s)} \bm{\hat{N}}(s)\, ds \, . \label{eq:quadrature_time_evolution}
\end{equation}
\noindent Equipped with the fact that the initial states and input noise have zero average, direct calculation shows that $\langle \bm{\hat{X}}(t) \rangle  =\bm{0}$ for all times.

The second moments of the system can be represented by the covariance matrix (CM), with components defined as $V_{i,j} = \frac{1}{2}\langle \hat{X}_i \hat{X}_j + \hat{X}_j \hat{X}_i\rangle$. Using Equation \eqref{eq:quadrature_time_evolution}, we see that the CM satisfies the Lyapunov equation
\begin{equation}
    \dot{V} = A V + VA^T + D \, , \label{eq:lyapunov}
\end{equation}
\noindent where $D_{l,k} \, \delta(t-t') \equiv \frac{1}{2}\langle \hat{N}_l(t) \hat{N}_k(t') + \hat{N}_k(t')\hat{N}_l(t) \rangle$, such that $D = \text{diag}\big(\kappa, \kappa, 0, 2\gamma_1(2\overbar{n}_{\mathrm{th}, 1} + 1), 0, 2\gamma_2(2\overbar{n}_{\mathrm{th}, 2} + 1), \ldots\big)$ is the \textit{drift} matrix. 
We can use the above dynamical equations to study entanglement within our system both in the closed and open regimes. 

\section{Unitary Entanglement Dynamics} \label{sec:unitary_entanglement}

It is expected that the unitary dynamics generated by \eqref{eq:CS_Hamiltonian_final} exhibits entanglement between the various mechanical and optical modes in the system. 
To gain some analytical insight into this entanglement generation consider the formal limit $ \omega_{j} \ll \Delta $. For $ N = 2 $ NPs the Hamiltonian reduces to, 
\begin{align}
    \H_{0}/\hbar &\approx  \Delta \a\dager\a 
    + g\,(\a\dager+\a)(\hat{x}_{1} + \hat{x}_{2}) \, , \label{eq:CS_Hamiltonian_unitary_example}
\end{align}
where we assume for simplicity that both NPs couple equally to the optical mode, $ g_{1} = g_{2} \equiv g $.
This simplified Hamiltonian can be exponentiated exactly using the same techniques employed in calculating the unitary evolution operator for a dispersive optomechanical system \cite{Brandao2020}.
The unitary time evolution operator for this case is $ U_{0}(t) = \exp (- i \H_{0} t /\hbar )$, which reads
\begin{align}
    \hat{U}_0(t) =  \me^{-i  \hat{a}^{\dagger}\hat{a} \Delta t} 
    \me^{(\hat{a}\eta(t)-\hat{a}^{\dagger}\eta(t)^*)(g/\Delta)(\hat{x}_1+\hat{x}_2)} \times \nonumber \\ 
    \times \me^{-i(g/\Delta)^{2}(\hat{x}_1+\hat{x}_2)^2\sin(\Delta\, t)} \me^{i g^{2} (\hat{x}_1-\hat{x}_2)^2 \Delta t}
\end{align}
where $ \eta(t) = 1-\me^{-i\Delta t} $. Note that the two exponential terms in the second line contain effective interactions among the NPs, given by products of the $ \hat{x}_1 , \hat{x}_2 $ terms. The presence of these optically mediated interactions lead to generation of entanglement between the NPs within this simplified approximation. 
Moreover, the the fact that the unitary operator is written in terms of periodic functions hints at entanglement death and revivals.

\begin{figure}[!ht]
    \centering
    \includegraphics[width=8.6cm]{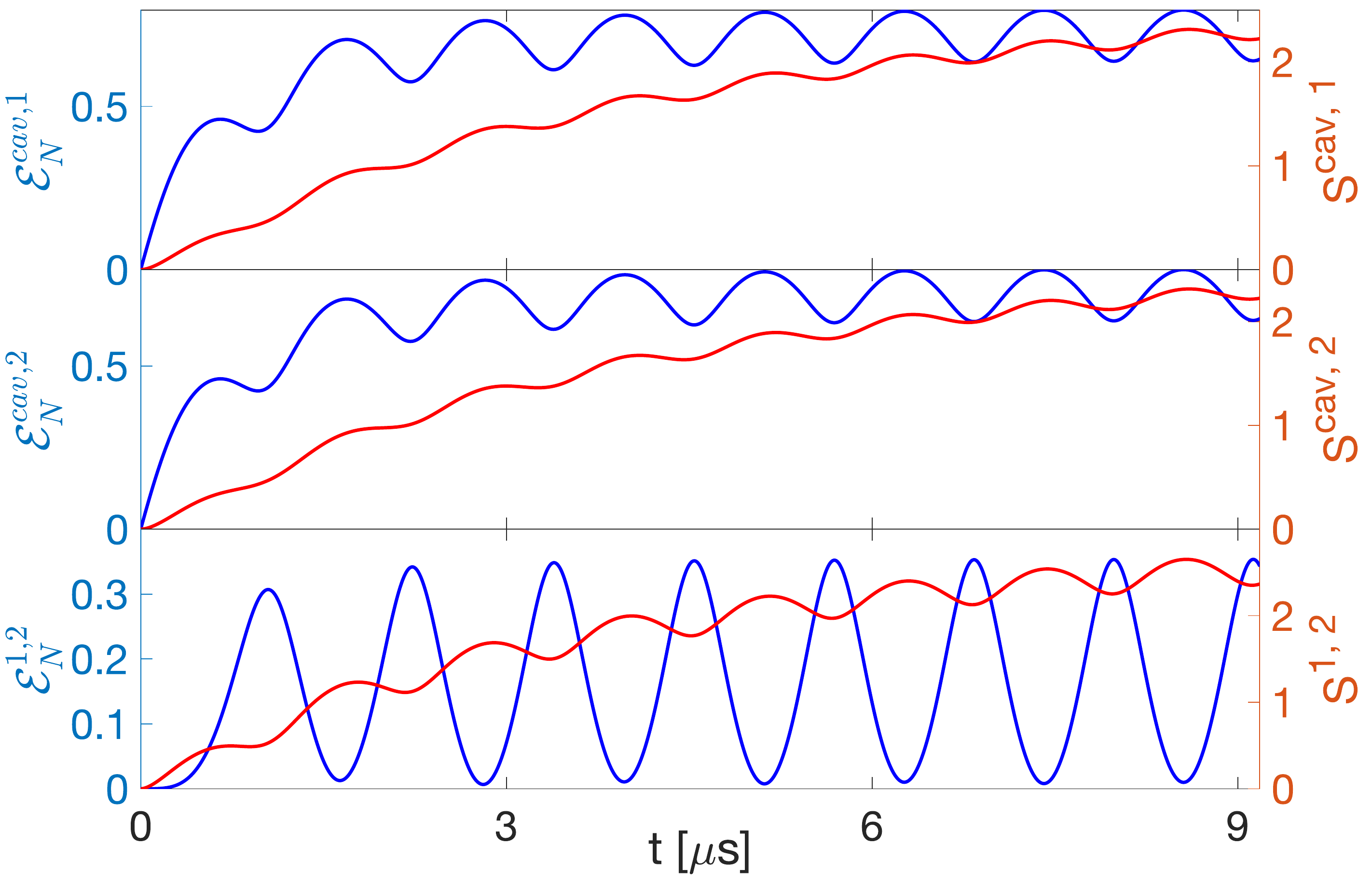}
    \caption{Simulation with $N=2$ identical levitated NPs initially in the vacuum state following a closed unitary dynamics. Time evolution of logarithmic negativity (blue) and von Neumann entropy (red) for each bipartition of the system. In this simulation, we considered $\kappa=0$ and $\gamma_j=0 \, \forall j$, making the system's dynamics closed.}
    \label{fig:unitary_time_evolution}
\end{figure}

In order to verify the entanglement generation in this regime, we numerically integrate the Lyapunov equation in the absence of noise and losses.
Figure \ref{fig:unitary_time_evolution} shows numerical plots of the Logarithmic negativity (LN) and von Neumann entropies for the various partitions of a system comprised of two NPs and one optical mode, all initially in the ground state. 
Note that both the LN and von Neumann entropy can be obtained by measurements of the covariance matrix \cite{Lloyd2012, Rakhubovsky2020} using a scheme analogous to the one described in \cite{Vitali2007}, based on homodyne detection. Note also that, provided the dynamics of the system is unitary and the initial state is pure, the emergence of a non-vanishing von Neumann entropy for partitions of the whole system signals entanglement.
The parameters used for this simulations are shown in Table \ref{tab:proposed_parameters}, except for $\kappa=0$ and $\gamma_j=0 \, \forall j$ to enforce unitary dynamics. 
From now on, unless explicitly stated otherwise, all NPs are taken to be identical and Table \ref{tab:proposed_parameters} dictates all the parameters considered in the simulations throughout the rest of this work; see Section \ref{sec:experimental_values}. 

We observe that in the unitary CS scenario cyclic entanglement birth and death are present, analogous to the dispersive optomechanical case \cite{Brandao2020}. Moreover, while the entropy of each bipartition is synchronized, the LN in the mechanical bipartition only achieves local maxima when the optomechanical LN are at their local minima. This points towards the idea that under certain circumstances entanglement can flow through different partitions of the system, in this case back and forth between the opto-mechanical and mechanical modes. 
It is also instructive to consider the optical field as an environment for the two NPs. Under this point of view, we can understand entanglement and entropy oscillations as a consequence of the non-Markovian nature of the subsystems' evolution.

\section{Entanglement in a noisy environment} \label{sec:open_entanglement_dynamics}

In any realistic experimental scenario the system under study is always interacting with its environment. For this reason it is important to study how the unitary entanglement dynamics is modified when the optomechanical system is placed in contact with uncontrolled external degrees of freedom. For high-vacuum environments ($p  < \SI{10^{-9}}{m\bar}$), it has been shown that CS mediated mechanical entanglement can resist photon scattering decoherence \cite{Hornberger2020} and in moderate vacuum ($p  \sim \SI{10^{-6}}{m\bar}$) steady state entanglement is only possible at low environment temperatures around $T\envj \approx \SI{15}{\kelvin}$ \cite{Chauhan2020}. In the latter regime, the crucial question of whether mechanical entanglement could exist for realistic environmental temperatures before the system achieves its steady state remains open.

\begin{figure}[!ht]
    \centering
    \includegraphics[width=8.6cm]{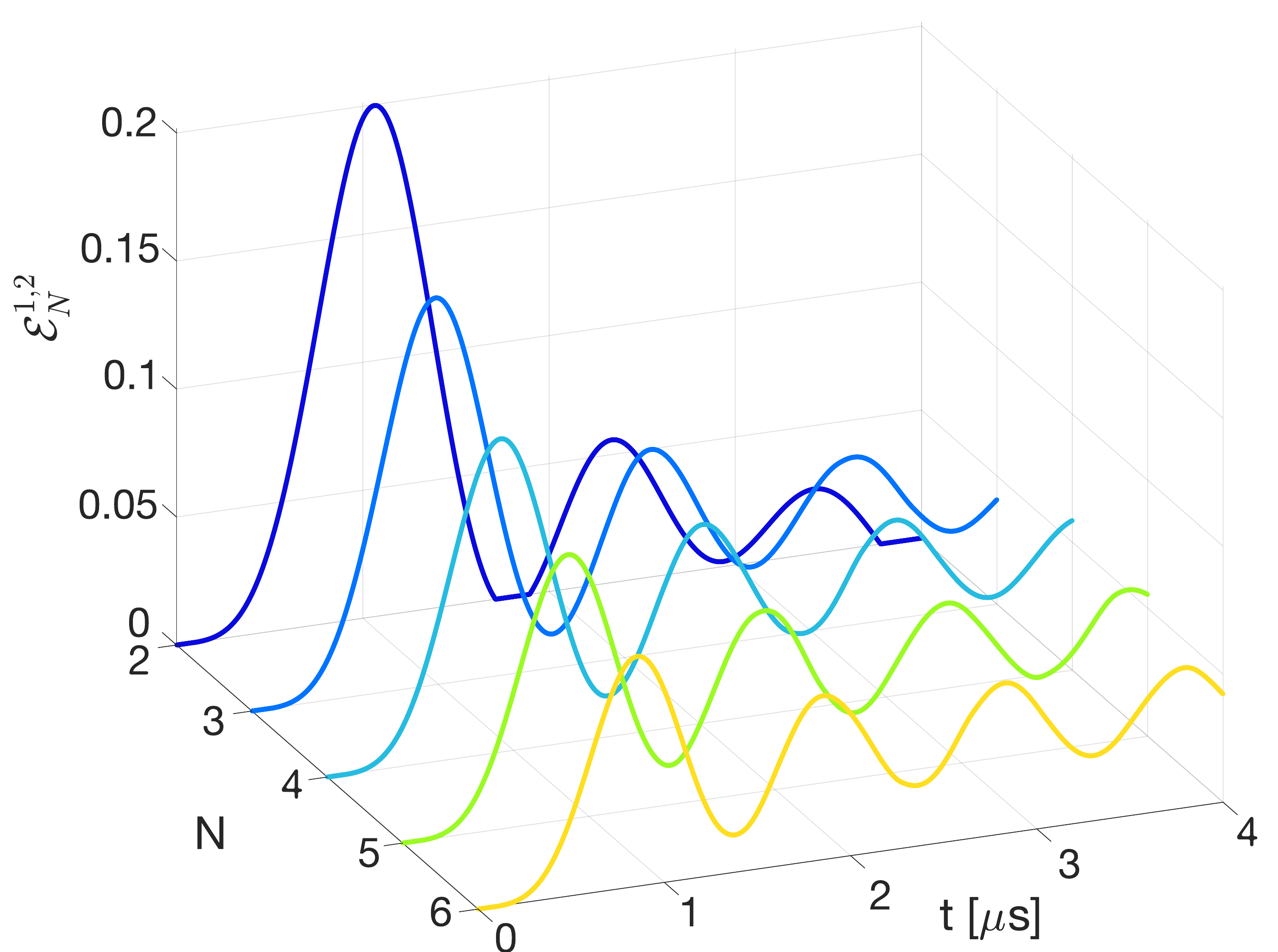}
    \caption{Time evolving LN between a pair of particles for an increasing number of particles sharing the same cavity. Every NP is considered to be in the ground state and every other parameter is shown in Table \ref{tab:proposed_parameters}.
    } 
    \label{fig:log_neg_many_particles_time}
\end{figure}

For a start, consider the more experimentally challenging scenario where each NP begins in the ground state.
This setting allows for entanglement of a large number of NPs. Figure \ref{fig:log_neg_many_particles_time} shows the time evolving LN as a function of the number of identical NPs in the cavity. We note that in this case the LN is symmetric over all possible mechanical bipartitions provided the particle parameters are identical, e.g. mass, coupling strength, bath temperature. 

Even in the presence of a room temperature Markovian environment, the oscillatory nature of the quantum correlations persists, revealing revivals of LN. Nonetheless, as the number of NPs in the cavity is increased, the amplitude of the LN curves decreases: the additional NPs act as an environment for the bipartite subsystem, i.e., the entanglement \textit{dilutes} over the system due to monogamy constraints. Although we discovered perseverance of entanglement for many particles systems, as $N$ increases, the system becomes unstable presenting no steady state nor stability in the numerical simulation for long times. Thus, Figure \ref{fig:log_neg_many_particles_time} shows the LN curves only up to $t\approx \SI{4}{\mu s}$.

\begin{figure*}[ht!]
    \centering
        \includegraphics[width = \textwidth]{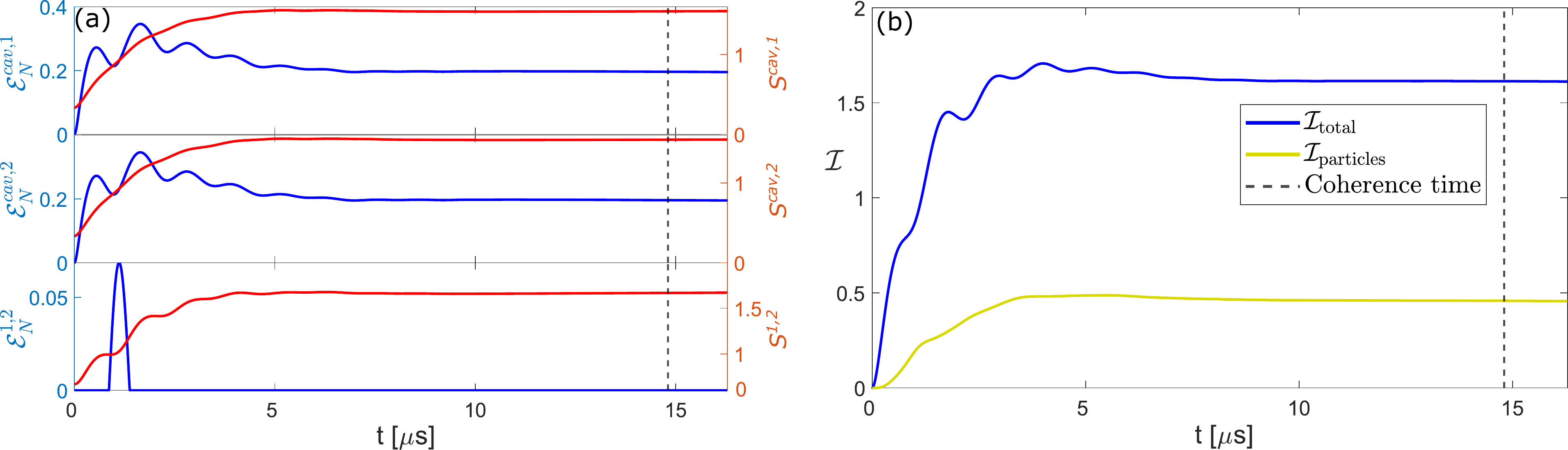}
    \caption{Simulation with $N=2$ identical levitated NPs initially in a thermal state with $0.1$ occupation number in contact with a room temperature environment, the parameters for these simulations are shown in Table \ref{tab:proposed_parameters}. (a) Time evolution of logarithmic negativity (blue) and von Neumann entropy (red) for each bipartition of the system. (b) Mutual information for the total system (yellow) and mechanical subsystem with only the levitated NPs (blue) evolving in time.}
    \label{fig:entanglement_and_entropy}
\end{figure*}

In the following we restrict our attention to a system comprised of $N=2$ identical levitated NPs, which allows us to study the system in a setting where entanglement generation is maximized.  
In Figure \ref{fig:entanglement_and_entropy}(a), we illustrate traces of the LN and entropy evolving in time for each possible bipartition of a system, where the particles are initially considered to be cooled to a thermal state at temperature $\SI{6.1}{\mu K}$, with occupation number $\overbar{n}_j=0.1$, close to the ones achieved in current experiments with optically levitated particles \cite{Delic2020a, Magrini2020}.
In contrast to the results of the previous simulations, we observe that quantum correlations between mechanical modes are only non-zero for a brief interval of time, while their oscillating nature is washed out by the initial non-zero occupation numbers and interactions with the environment.
Note, however, that the opto-mechanical entanglement persists over some oscillations before dying out in the steady state 

\noindent at long times (not shown). 

Note also that the non-Markovian features of the evolution such as oscillations of entropy, albeit present, are strongly attenuated due to the nature of environmental interactions, since the environment is traced out and treated effectively as damping and stochastic forces in a Markovian approximation. 

We can speak more broadly about the correlations within the system by employing the mutual information, a measure of the total classical and quantum correlations \cite{Groisman2005}; see Appendix \ref{appendix:gaussian_toolbox} for details on the definition. In Figure \ref{fig:entanglement_and_entropy}(b), we plot the time evolving mutual information for the total system, $\mathcal{I}_{\rm total}$, and for the reduced system composed solely of the NPs, $\mathcal{I}_{\rm particles}$. The parties are initially uncorrelated as expected from the form of the separable states at $ t = 0 $. It later becomes correlated during the system's coherence time. In the reduced mechanical bipartition, CS-mediated correlations are generated before quantum entanglement comes into play and persists after entanglement death.
As a final remark, we have solved the Lyapunov equation for the steady state, given by $AV + VA^T + D = 0$, and found that $\mathcal{E}_N^{j,k} = 0 \, \forall \, j,k$, which is consistent with previous results in the literature \cite{Chauhan2020}. 
However, the steady state displays a non-zero total mutual information of $\mathcal{I}_{\rm total} \simeq 16.3$ and $\mathcal{I}_{\rm particles} \simeq 15.1$ (not shown in Figure \ref{fig:entanglement_and_entropy}(b)), meaning that although the LN cannot detect steady state entanglement, general correlations among subsystems due to the CS interaction are present. 
To confirm these results, we ran the numerical simulation for long times, and indeed found that the mutual information slowly ramps up to the predicted value $\approx 15$, for a time much larger than the coherence time of the system.

So far, our focus has been centered on how coherent scattering-mediated entanglement appears and evolves in time. We now shift our attention to the LN's dependency on experimentally controlled parameters, comparing with current state of the art experiments with optically levitated nanospheres. An in depth discussion of the experimental feasibility of the parameters used in the simulations is presented in Section \ref{sec:experimental_values} and Appendix \ref{appendix:experimental_params}.


\begin{figure}[!ht]
    \centering
    \includegraphics[width=8.6cm]{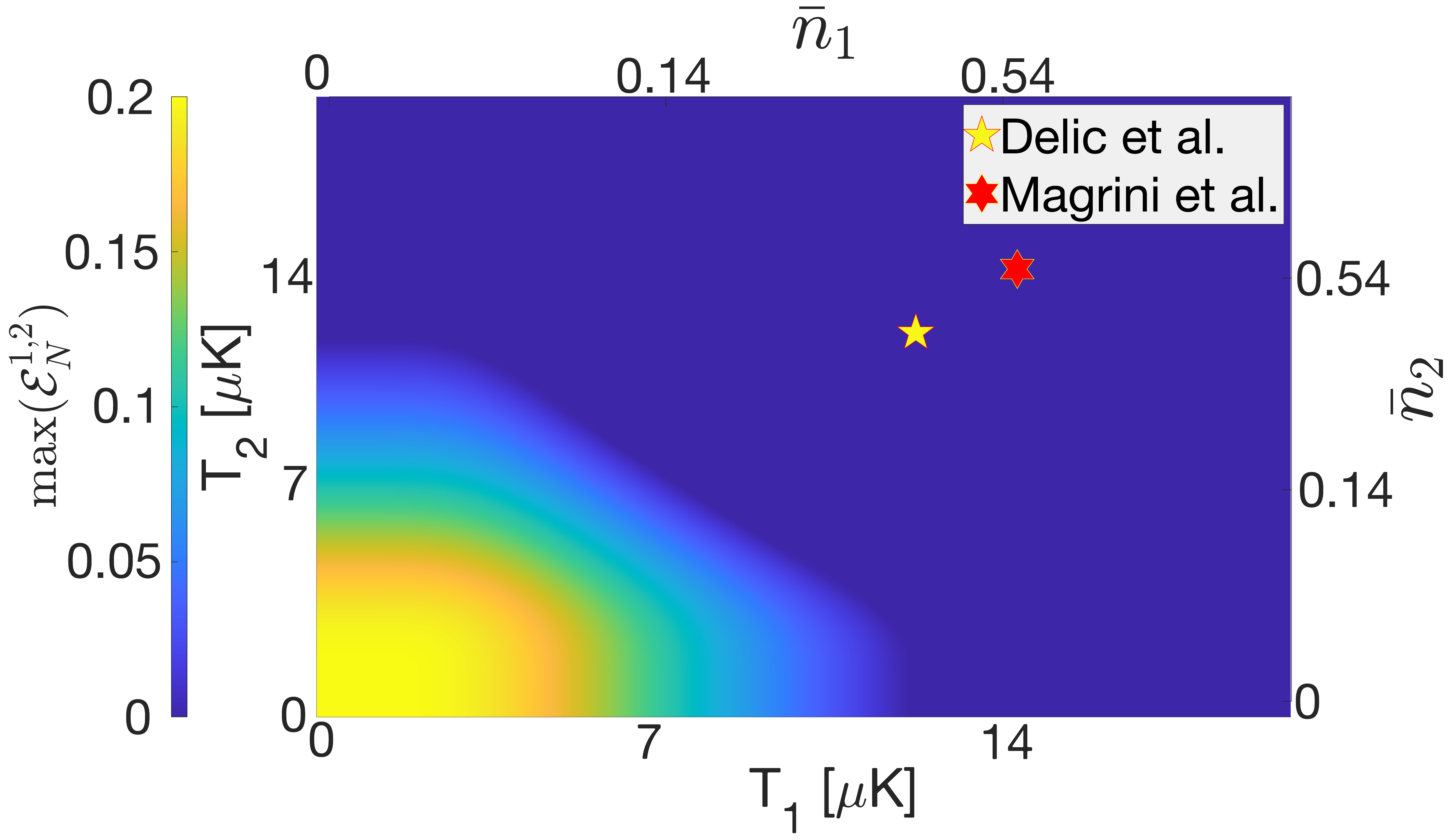}
    \caption{Maximum of the logarithmic negativity within the system's coherence time for two particles as a function of their initial temperatures.
    Notice that we need highly cooled particles to generate mechanical entanglement. 
    \label{fig:log_neg_temperature}} 
\end{figure}


\begin{figure}[!ht]
    \centering
    \includegraphics[width=8.6cm]{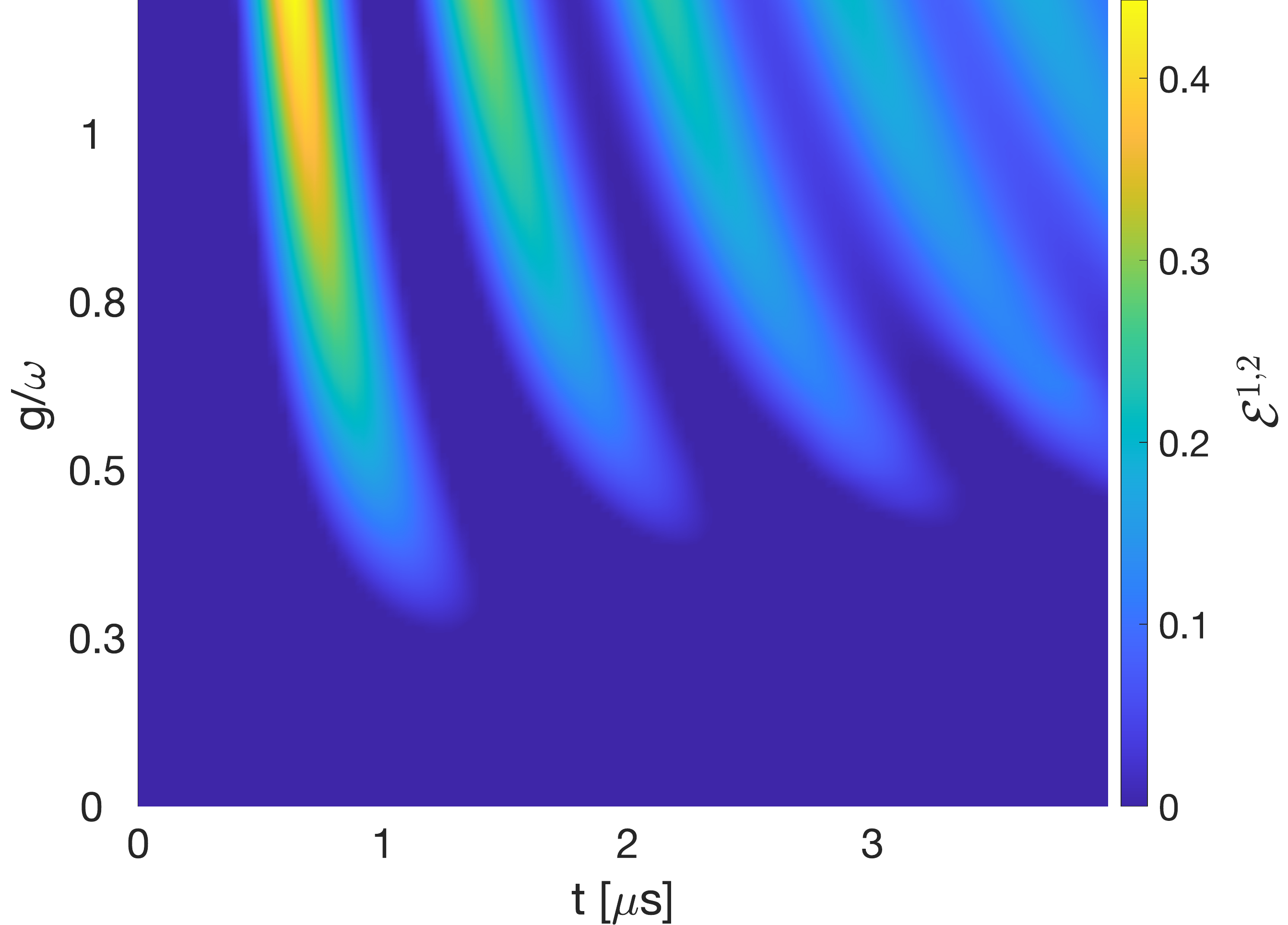}
    \caption{Time evolving logarithmic negativity for different coupling strengths, $g_j = g $ for all $ j$. Here we considered every NP starting in a thermal state at  $T_j \sim \SI{6.1}{\mu\kelvin}$ (occupation number $\overbar{n}_{0,j} = 0.1$). Observe that in the high coupling regime, $g \sim \omega$, entanglement birth, death and revivals are present.}
    \label{fig:log_neg_coupling_time}
\end{figure} 


\begin{figure*}[t!]
    \centering
    \includegraphics[width = 0.8\textwidth]{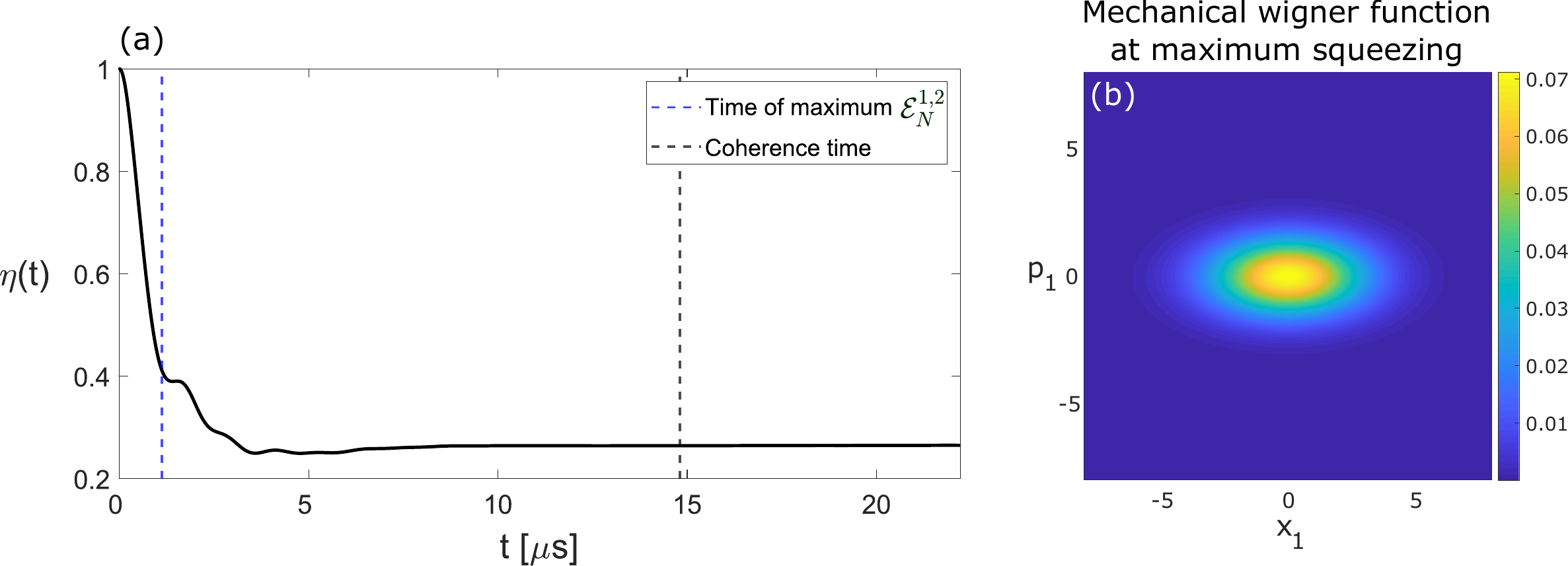}
    \caption{Time evolution of the mechanical squeezing for $N=2$ identical levitated NPs using the parameters shown in Table \ref{tab:proposed_parameters}. As the particles are identical, they posses identical Wigner functions.
    (a) Time evolving squeezing parameter $\eta(t)$ for a single particle. Black dashed lines indicate coherence time and blue dashed line indicates the time of maximum mechanical entanglement. As the particles are initially in a thermal state the squeezing degree starts at $\eta(0) = 1$ (no squeezing) and achieves a global minimum of $\eta(t\simeq\SI{5.7}{\mu s}) \simeq 0.23$ before squeezing vanishes once again in the steady state (not shown). (b) Wigner function for each nanosphere at the time of maximum squeezing.}
    \label{fig:Wigner}
\end{figure*}

Figure \ref{fig:log_neg_temperature} shows the maximum of the LN between the two NPs within the coherence time of the system, $\tau \simeq \SI{14.8}{\mu \second}$ as quantified in Appendix \ref{appendix:experimental_params}, for different values of initial particle's temperatures and occupation number in the presence of an environment at room temperature ($T\envj=\SI{300}{K} $). 
We can immediately see that mechanical entanglement would only occur for extremely pre-cooled particles. For initial occupations numbers bigger $\overbar{n}_j > 0.16$, the LN vanishes. Nonetheless, we note that such occupations numbers are less than an order of magnitude away from the occupations achieved in current ground state cooling experiments with such nanospheres \cite{Delic2019, Magrini2020}, respectively marked with a yellow five-pointed star and a red six-pointed star on the plot on the line with equal initial temperatures as if two particles cooled through their techniques were put to indirectly interact in a CS scheme. Although further cooling is necessary, it ought to be experimentally feasible to observe mechanical entanglement in this scheme.  


This difficulty could be partially circumvented by increasing the CS coupling. In Figure \ref{fig:log_neg_coupling_time}, we show the time-dependent LN as a function of the optomechanical coupling strength for an initial NP temperature of $T_j \sim \SI{6.1}{\mu\kelvin}$ (occupation number $\overbar{n}_{0,j} = 0.1$).
We readily see that entanglement generation depends significantly on the optomechanical coupling strength. 
Current state-of-the-art CS implementations achieve $g_j / \omega_j \sim 0.2$ \cite{Delic2020a}, below the region in which entanglement and its revivals are present. 
We note that couplings as high as $\SI{2\pi \times 110}{\kHz}$ could be achieved by properly adjusting the NP radius and tweezers' power and waist, see Appendix \ref{appendix:experimental_params}. In such regime, where $g_j / \omega_j \sim 0.36$, entanglement starts to appear in the system. This can be promptly seen by noticing that a horizontal cross cut of Figure \ref{fig:log_neg_coupling_time} at $g_j / \omega_j \sim 0.36$ corresponds to the plot in Figure \ref{fig:entanglement_and_entropy}(a). If it is possible to enhance even further the coupling strength, we start to retrieve the cyclic dynamics of birth, death and revivals of entanglement. For higher values of optomechanical coupling, we enter the strong coupling regime $g_j \geq \omega_j$, where a plateau of entanglement occurs and entanglement survives for more than brief periods of time.


As a concluding remark, we note that another interesting feature of the CS interaction is the generation of mechanical squeezing during the system's time evolution. Generation of squeezing using the CS Hamiltonian for a single particle has been shown in \cite{Radim2020}. We are able to show that squeezing is also generated if more than one particle is present in the cavity through the so-called squeezing parameter, $\eta$ \cite{Radim2020}. It is defined as the ratio of the variance of the squeezed and the antisqueezed quadratures such that squeezing can be inferred if $\eta < 1$, see Appendix \ref{appendix:decoherence} for more details. As an example, consider the case of two NPs in the cavity initially in a thermal state, $\eta(0)=1$. Figure \ref{fig:Wigner}(a) displays the time evolving squeezing parameter for one of the particles alongside Figure \ref{fig:Wigner}(b) with its Wigner function at the time of maximum squeezing, $t \simeq \SI{5.7}{\mu s}$. As we consider identical NPs, their Wigner functions are also identical and only a single one is shown. 
We observe the generation of squeezing through the appearance of elliptical shapes of the Wigner distribution and the calculated lower than one $\eta(t)$. For long times, however, the squeezing parameter slowly increases until all squeezing vanishes on the steady state (not shown in Figure \ref{fig:Wigner}(a), but verified numerically with a simulation for times much longer than the coherence time). We also comment that emergence of squeezing in multi-particle CS could find interesting applications in quantum metrology of tiny forces \cite{Weiss2020}. 


\section{Experimental parameters} \label{sec:experimental_values}

The parameters used in our simulations were adapted from Ref. \cite{Delic2020a}. There, a silica nanosphere (density of SiO${}_2$: $\SI{2200}{kg/m^3}$ \cite{Gonzalez-Ballestero2019}) was trapped by an OT within a high-finesse Fabry-Pérot cavity mounted in a vacuum chamber and cavity cooling through CS was performed to cool the CM of the NP into its motional ground state. The parameters values of this experimental realization are presented in Table \ref{tab:experimental_values_delic}.

\begin{table}[!ht]
\centering
\caption{Experimental parameters reported in Ref. \cite{Delic2020a}}
\label{tab:experimental_values_delic}
\begin{tabular}{lcc}
\hline
\hline 
Parameter & Unit & Value\\
    \midrule
Tweezer power $P\tw$ & $\SI{}{\milli\watt}$ & $400$ \\
Tweezer wavelength $\lambda\tw$ & $\SI{}{\nano\meter}$ & $1064$ \\
Tweezer waist (x axis)  & $\SI{}{\mu\meter}$ & $0.67$ \\
Tweezer waist (y axis)  & $\SI{}{\mu\meter}$ & $0.77$ \\
Trapping frequency $\omega_x$ & $\SI{}{\kilo\hertz}$ & $2\pi\times305$\\
Cavity finesse $\mathcal{F}$ & --- & $73,000$ \\
Cavity linewidth $\kappa$  & $\SI{}{\kilo\hertz}$ & $2\pi\times193$ \\
Cavity length $L$ &  $\SI{}{\milli\meter}$ & $10.7$\\
Cavity waist $w_{0,\mathrm{c}}$ &  $\SI{}{\micro\meter}$ & $41.1$\\
Cavity-tweezer detuning $\Delta$ & $\SI{}{\kilo\hertz}$ & $2\pi\times315$ \\
Particle mass $m$ & $\SI{}{\femto\gram}$ & $2.83$ \\
Particle radius $R$ & $\SI{}{\nano\meter}$ & $71.5$ \\
Particle zero point fluctuation $x_{\rm ZPF}$ & $\SI{}{\pico\meter}$ & $3.1$ \\
Vacuum chamber pressure $p_{\rm gas}$ & $\SI{}{m\bar}$ & $10^{-6}$ \\
Environmental temperature $T_{\rm gas}$ & $\SI{}{K}$ & $300$ \\
\hline\hline
\end{tabular}
\end{table}


When considering two NPs with these parameters and initial occupation number of $\overbar{n}_{0,j} = 0.43$, achieved in Ref. \cite{Delic2019}, we observed zero logarithmic negativity in our simulations. In order to be able to observe some evidence of mechanical entanglement in this system, we studied the dependence of the coherence time and coupling strength with the experimentally parameters above in Appendix \ref{appendix:experimental_params}. We discovered a parameter region that allows for such quantum correlations to occur with lower decoherence rates and greater coupling strength. Thus, we propose slight adaptations to he tweezer's waist and power, and particle size, as detailed in Appendix \ref{appendix:experimental_params}. Table \ref{tab:proposed_parameters} shows the resulting parameters used in the simulation of this work.

\begin{table}[!ht]
\centering
\caption{Parameter values used in this work}
\label{tab:proposed_parameters}
\begin{tabular}{lcc}
\hline
\hline 
Parameter & Unit & Value\\
    \midrule
Mechanical natural frequency $\omega$            & $\SI{}{\kilo\hertz}$  & $2\pi\times305.26$ \\
Cavity-tweezer detuning $\Delta$                 & $\SI{}{\kilo\hertz}$  & $2\pi\times315.0$ \\
CS coupling strength $g$                         & $\SI{}{\kilo\hertz}$  & $2\pi\times109.8$ \\
Cavity linewidth $\kappa$                        & $\SI{}{\kilo\hertz}$  & $2\pi\times193.0$ \\
Damping coefficient $\gamma$                     & $\SI{}{\mu\hertz}$ & $2\pi\times5.88$ \\
Particle initial temperature $T_0$               & $\SI{}{\mu\kelvin}$   & $6.1$            \\
Particle initial occupation number               & ---                     & 0.1 \\
Environmental gas temperature $T_{\rm gas}\quad$ & $\SI{}{\kelvin}$      & $300$             \\
\hline\hline
\end{tabular}
\end{table}

Unless explicitly stated otherwise, these are the values considered in the simulations throughout this work with all NPs identical to each other.

\section{Conclusion}

We have extended the CS Hamiltonian to an arbitrary number of particles, describing $ N $ mechanical oscillators interacting with a single cavity mode. The resulting unitary dynamics has been shown to generate quantum correlations in every bipartition of the system following cycles of entanglement birth and death. Moreover, within each revival, the Hamiltonian appears to steam a flow of quantum correlations between the bipartitions, with mechanical entanglement maximized exactly when opto-mechanical entanglement is minimized.

In realistic experimental conditions, one has to consider the environmental effects on the system. Following an open quantum dynamics, we have show that entanglement generation can still persist even in room temperature environments for some time within experimentally reasonable parameters. Although in the steady state these quantum correlations die out \cite{Chauhan2020}, we have shown that general correlations are still present in the system.

We studied the dependence of mechanical entanglement on experimentally controlled parameters. We discovered that increasing the number of NPs interacting with the optical field \textit{dilutes} the entanglement over the complete system and is detrimental to the creation of bipartite quantum correlations. Such correlations are expected to only appear when the particles are extremely cooled down, below the minimum occupancy achieved with current state-of-the-art technology \cite{Delic2020a, Magrini2020}. The CS coupling strength also plays a significant role in entanglement generation, and revivals of entanglement only come into play in the regime where $g_j > 0.3 \ \omega_j$. Squeezing is also generated by the many-particle CS Hamiltonian. 
In summary, the coherent scattering interaction propels optomechanics to the domain of complex quantum systems within realistic scenarios. It will be interesting to see what new experiments are enabled by this many-body mesoscopic quantum toolbox.

\section*{Acknowledgements}
We thank Bruno Suassuna and Bruno Melo for helpful discussions. This work was financed in part by the Serrapilheira Institute (grant number Serra-1709-21072), by Coordenaç\~ao de Aperfei\c{c}oamento de Pessoal de N\'ivel Superior - Brasil (CAPES) - Finance Code 001, by Conselho Nacional de Desenvolvimento Cient\'ifico e Tecnol\'ogico (CNPq). I. B. thanks the support received by the FAPERJ Scholarship No. E-26/200.270/2020 and CNPq Scholarship No. 140279/2021-0. T.G. thanks the  support received by the FAPERJ Scholarship No. E-26/202.830/2019. D. T. thanks the support received by the CNPq Scholarship No. 132606/2020-8.

\bibliography{main.bib}

\begin{thebibliography}{59}%
\makeatletter
\providecommand \@ifxundefined [1]{%
 \@ifx{#1\undefined}
}%
\providecommand \@ifnum [1]{%
 \ifnum #1\expandafter \@firstoftwo
 \else \expandafter \@secondoftwo
 \fi
}%
\providecommand \@ifx [1]{%
 \ifx #1\expandafter \@firstoftwo
 \else \expandafter \@secondoftwo
 \fi
}%
\providecommand \natexlab [1]{#1}%
\providecommand \enquote  [1]{``#1''}%
\providecommand \bibnamefont  [1]{#1}%
\providecommand \bibfnamefont [1]{#1}%
\providecommand \citenamefont [1]{#1}%
\providecommand \href@noop [0]{\@secondoftwo}%
\providecommand \href [0]{\begingroup \@sanitize@url \@href}%
\providecommand \@href[1]{\@@startlink{#1}\@@href}%
\providecommand \@@href[1]{\endgroup#1\@@endlink}%
\providecommand \@sanitize@url [0]{\catcode `\\12\catcode `\$12\catcode
  `\&12\catcode `\#12\catcode `\^12\catcode `\_12\catcode `\%12\relax}%
\providecommand \@@startlink[1]{}%
\providecommand \@@endlink[0]{}%
\providecommand \url  [0]{\begingroup\@sanitize@url \@url }%
\providecommand \@url [1]{\endgroup\@href {#1}{\urlprefix }}%
\providecommand \urlprefix  [0]{URL }%
\providecommand \Eprint [0]{\href }%
\providecommand \doibase [0]{https://doi.org/}%
\providecommand \selectlanguage [0]{\@gobble}%
\providecommand \bibinfo  [0]{\@secondoftwo}%
\providecommand \bibfield  [0]{\@secondoftwo}%
\providecommand \translation [1]{[#1]}%
\providecommand \BibitemOpen [0]{}%
\providecommand \bibitemStop [0]{}%
\providecommand \bibitemNoStop [0]{.\EOS\space}%
\providecommand \EOS [0]{\spacefactor3000\relax}%
\providecommand \BibitemShut  [1]{\csname bibitem#1\endcsname}%
\let\auto@bib@innerbib\@empty
\bibitem [{\citenamefont {Bohnet}\ \emph {et~al.}(2016)\citenamefont {Bohnet},
  \citenamefont {Sawyer}, \citenamefont {Britton}, \citenamefont {Wall},
  \citenamefont {Rey}, \citenamefont {Foss-Feig},\ and\ \citenamefont
  {Bollinger}}]{Bohnet2016}%
  \BibitemOpen
  \bibfield  {author} {\bibinfo {author} {\bibfnamefont {J.~G.}\ \bibnamefont
  {Bohnet}}, \bibinfo {author} {\bibfnamefont {B.~C.}\ \bibnamefont {Sawyer}},
  \bibinfo {author} {\bibfnamefont {J.~W.}\ \bibnamefont {Britton}}, \bibinfo
  {author} {\bibfnamefont {M.~L.}\ \bibnamefont {Wall}}, \bibinfo {author}
  {\bibfnamefont {A.~M.}\ \bibnamefont {Rey}}, \bibinfo {author} {\bibfnamefont
  {M.}~\bibnamefont {Foss-Feig}},\ and\ \bibinfo {author} {\bibfnamefont
  {J.~J.}\ \bibnamefont {Bollinger}},\ }\href
  {https://doi.org/10.1126/science.aad9958} {\bibfield  {journal} {\bibinfo
  {journal} {Science}\ }\textbf {\bibinfo {volume} {352}},\ \bibinfo {pages}
  {1297} (\bibinfo {year} {2016})}\BibitemShut {NoStop}%
\bibitem [{\citenamefont {Xu}\ \emph {et~al.}(2019)\citenamefont {Xu},
  \citenamefont {Jaffe}, \citenamefont {Panda}, \citenamefont {Kristensen},
  \citenamefont {Clark},\ and\ \citenamefont {M\"{u}ller}}]{Xu2019}%
  \BibitemOpen
  \bibfield  {author} {\bibinfo {author} {\bibfnamefont {V.}~\bibnamefont
  {Xu}}, \bibinfo {author} {\bibfnamefont {M.}~\bibnamefont {Jaffe}}, \bibinfo
  {author} {\bibfnamefont {C.~D.}\ \bibnamefont {Panda}}, \bibinfo {author}
  {\bibfnamefont {S.~L.}\ \bibnamefont {Kristensen}}, \bibinfo {author}
  {\bibfnamefont {L.~W.}\ \bibnamefont {Clark}},\ and\ \bibinfo {author}
  {\bibfnamefont {H.}~\bibnamefont {M\"{u}ller}},\ }\href
  {https://doi.org/10.1126/science.aay6428} {\bibfield  {journal} {\bibinfo
  {journal} {Science}\ }\textbf {\bibinfo {volume} {366}},\ \bibinfo {pages}
  {745} (\bibinfo {year} {2019})}\BibitemShut {NoStop}%
\bibitem [{\citenamefont {Berrada}\ \emph {et~al.}(2013)\citenamefont
  {Berrada}, \citenamefont {van Frank}, \citenamefont {B\"{u}cker},
  \citenamefont {Schumm}, \citenamefont {Schaff},\ and\ \citenamefont
  {Schmiedmayer}}]{Berrada2013}%
  \BibitemOpen
  \bibfield  {author} {\bibinfo {author} {\bibfnamefont {T.}~\bibnamefont
  {Berrada}}, \bibinfo {author} {\bibfnamefont {S.}~\bibnamefont {van Frank}},
  \bibinfo {author} {\bibfnamefont {R.}~\bibnamefont {B\"{u}cker}}, \bibinfo
  {author} {\bibfnamefont {T.}~\bibnamefont {Schumm}}, \bibinfo {author}
  {\bibfnamefont {J.-F.}\ \bibnamefont {Schaff}},\ and\ \bibinfo {author}
  {\bibfnamefont {J.}~\bibnamefont {Schmiedmayer}},\ }\bibfield  {journal}
  {\bibinfo  {journal} {Nature Communications}\ }\textbf {\bibinfo {volume}
  {4}},\ \href {https://doi.org/10.1038/ncomms3077} {10.1038/ncomms3077}
  (\bibinfo {year} {2013})\BibitemShut {NoStop}%
\bibitem [{\citenamefont {Huo}\ and\ \citenamefont {Long}(2008)}]{Huo2008}%
  \BibitemOpen
  \bibfield  {author} {\bibinfo {author} {\bibfnamefont {W.~Y.}\ \bibnamefont
  {Huo}}\ and\ \bibinfo {author} {\bibfnamefont {G.~L.}\ \bibnamefont {Long}},\
  }\href {https://doi.org/10.1063/1.2904700} {\bibfield  {journal} {\bibinfo
  {journal} {Applied Physics Letters}\ }\textbf {\bibinfo {volume} {92}},\
  \bibinfo {pages} {133102} (\bibinfo {year} {2008})}\BibitemShut {NoStop}%
\bibitem [{\citenamefont {Etaki}\ \emph {et~al.}(2008)\citenamefont {Etaki},
  \citenamefont {Poot}, \citenamefont {Mahboob}, \citenamefont {Onomitsu},
  \citenamefont {Yamaguchi},\ and\ \citenamefont {van~der Zant}}]{Etaki2008}%
  \BibitemOpen
  \bibfield  {author} {\bibinfo {author} {\bibfnamefont {S.}~\bibnamefont
  {Etaki}}, \bibinfo {author} {\bibfnamefont {M.}~\bibnamefont {Poot}},
  \bibinfo {author} {\bibfnamefont {I.}~\bibnamefont {Mahboob}}, \bibinfo
  {author} {\bibfnamefont {K.}~\bibnamefont {Onomitsu}}, \bibinfo {author}
  {\bibfnamefont {H.}~\bibnamefont {Yamaguchi}},\ and\ \bibinfo {author}
  {\bibfnamefont {H.~S.~J.}\ \bibnamefont {van~der Zant}},\ }\href
  {https://doi.org/10.1038/nphys1057} {\bibfield  {journal} {\bibinfo
  {journal} {Nature Physics}\ }\textbf {\bibinfo {volume} {4}},\ \bibinfo
  {pages} {785} (\bibinfo {year} {2008})}\BibitemShut {NoStop}%
\bibitem [{\citenamefont {O'Connell}\ \emph {et~al.}(2010)\citenamefont
  {O'Connell}, \citenamefont {Hofheinz}, \citenamefont {Ansmann}, \citenamefont
  {Bialczak}, \citenamefont {Lenander}, \citenamefont {Lucero}, \citenamefont
  {Neeley}, \citenamefont {Sank}, \citenamefont {Wang}, \citenamefont {Weides},
  \citenamefont {Wenner}, \citenamefont {Martinis},\ and\ \citenamefont
  {Cleland}}]{OConnell2010}%
  \BibitemOpen
  \bibfield  {author} {\bibinfo {author} {\bibfnamefont {A.~D.}\ \bibnamefont
  {O'Connell}}, \bibinfo {author} {\bibfnamefont {M.}~\bibnamefont {Hofheinz}},
  \bibinfo {author} {\bibfnamefont {M.}~\bibnamefont {Ansmann}}, \bibinfo
  {author} {\bibfnamefont {R.~C.}\ \bibnamefont {Bialczak}}, \bibinfo {author}
  {\bibfnamefont {M.}~\bibnamefont {Lenander}}, \bibinfo {author}
  {\bibfnamefont {E.}~\bibnamefont {Lucero}}, \bibinfo {author} {\bibfnamefont
  {M.}~\bibnamefont {Neeley}}, \bibinfo {author} {\bibfnamefont
  {D.}~\bibnamefont {Sank}}, \bibinfo {author} {\bibfnamefont {H.}~\bibnamefont
  {Wang}}, \bibinfo {author} {\bibfnamefont {M.}~\bibnamefont {Weides}},
  \bibinfo {author} {\bibfnamefont {J.}~\bibnamefont {Wenner}}, \bibinfo
  {author} {\bibfnamefont {J.~M.}\ \bibnamefont {Martinis}},\ and\ \bibinfo
  {author} {\bibfnamefont {A.~N.}\ \bibnamefont {Cleland}},\ }\href
  {https://doi.org/10.1038/nature08967} {\bibfield  {journal} {\bibinfo
  {journal} {Nature}\ }\textbf {\bibinfo {volume} {464}},\ \bibinfo {pages}
  {697} (\bibinfo {year} {2010})}\BibitemShut {NoStop}%
\bibitem [{\citenamefont {Chan}\ \emph {et~al.}(2011)\citenamefont {Chan},
  \citenamefont {Alegre}, \citenamefont {Safavi-Naeini}, \citenamefont {Hill},
  \citenamefont {Krause}, \citenamefont {Gr\"{o}blacher}, \citenamefont
  {Aspelmeyer},\ and\ \citenamefont {Painter}}]{Chan2011}%
  \BibitemOpen
  \bibfield  {author} {\bibinfo {author} {\bibfnamefont {J.}~\bibnamefont
  {Chan}}, \bibinfo {author} {\bibfnamefont {T.~P.~M.}\ \bibnamefont {Alegre}},
  \bibinfo {author} {\bibfnamefont {A.~H.}\ \bibnamefont {Safavi-Naeini}},
  \bibinfo {author} {\bibfnamefont {J.~T.}\ \bibnamefont {Hill}}, \bibinfo
  {author} {\bibfnamefont {A.}~\bibnamefont {Krause}}, \bibinfo {author}
  {\bibfnamefont {S.}~\bibnamefont {Gr\"{o}blacher}}, \bibinfo {author}
  {\bibfnamefont {M.}~\bibnamefont {Aspelmeyer}},\ and\ \bibinfo {author}
  {\bibfnamefont {O.}~\bibnamefont {Painter}},\ }\href
  {https://doi.org/10.1038/nature10461} {\bibfield  {journal} {\bibinfo
  {journal} {Nature}\ }\textbf {\bibinfo {volume} {478}},\ \bibinfo {pages}
  {89} (\bibinfo {year} {2011})}\BibitemShut {NoStop}%
\bibitem [{\citenamefont {Deli{\'{c}}}\ \emph
  {et~al.}(2020{\natexlab{a}})\citenamefont {Deli{\'{c}}}, \citenamefont
  {Reisenbauer}, \citenamefont {Dare}, \citenamefont {Grass}, \citenamefont
  {Vuleti{\'{c}}}, \citenamefont {Kiesel},\ and\ \citenamefont
  {Aspelmeyer}}]{Delic2020a}%
  \BibitemOpen
  \bibfield  {author} {\bibinfo {author} {\bibfnamefont {U.}~\bibnamefont
  {Deli{\'{c}}}}, \bibinfo {author} {\bibfnamefont {M.}~\bibnamefont
  {Reisenbauer}}, \bibinfo {author} {\bibfnamefont {K.}~\bibnamefont {Dare}},
  \bibinfo {author} {\bibfnamefont {D.}~\bibnamefont {Grass}}, \bibinfo
  {author} {\bibfnamefont {V.}~\bibnamefont {Vuleti{\'{c}}}}, \bibinfo {author}
  {\bibfnamefont {N.}~\bibnamefont {Kiesel}},\ and\ \bibinfo {author}
  {\bibfnamefont {M.}~\bibnamefont {Aspelmeyer}},\ }\href
  {https://doi.org/10.1126/science.aba3993} {\bibfield  {journal} {\bibinfo
  {journal} {Science}\ }\textbf {\bibinfo {volume} {367}},\ \bibinfo {pages}
  {892} (\bibinfo {year} {2020}{\natexlab{a}})}\BibitemShut {NoStop}%
\bibitem [{\citenamefont {Magrini}\ \emph {et~al.}(2021)\citenamefont
  {Magrini}, \citenamefont {Rosenzweig}, \citenamefont {Bach}, \citenamefont
  {Deutschmann-Olek}, \citenamefont {Hofer}, \citenamefont {Hong},
  \citenamefont {Kiesel}, \citenamefont {Kugi},\ and\ \citenamefont
  {Aspelmeyer}}]{Magrini2020}%
  \BibitemOpen
  \bibfield  {author} {\bibinfo {author} {\bibfnamefont {L.}~\bibnamefont
  {Magrini}}, \bibinfo {author} {\bibfnamefont {P.}~\bibnamefont {Rosenzweig}},
  \bibinfo {author} {\bibfnamefont {C.}~\bibnamefont {Bach}}, \bibinfo {author}
  {\bibfnamefont {A.}~\bibnamefont {Deutschmann-Olek}}, \bibinfo {author}
  {\bibfnamefont {S.~G.}\ \bibnamefont {Hofer}}, \bibinfo {author}
  {\bibfnamefont {S.}~\bibnamefont {Hong}}, \bibinfo {author} {\bibfnamefont
  {N.}~\bibnamefont {Kiesel}}, \bibinfo {author} {\bibfnamefont
  {A.}~\bibnamefont {Kugi}},\ and\ \bibinfo {author} {\bibfnamefont
  {M.}~\bibnamefont {Aspelmeyer}},\ }\href@noop {} {\bibinfo {title} {Real-time
  optimal quantum control of mechanical motion at room temperature}} (\bibinfo
  {year} {2021}),\ \Eprint {https://arxiv.org/abs/2012.15188} {arXiv:2012.15188
  [quant-ph]} \BibitemShut {NoStop}%
\bibitem [{\citenamefont {Kuhn}\ \emph {et~al.}(2017)\citenamefont {Kuhn},
  \citenamefont {Kosloff}, \citenamefont {Stickler}, \citenamefont {Patolsky},
  \citenamefont {Hornberger}, \citenamefont {Arndt},\ and\ \citenamefont
  {Millen}}]{Kuhn2017}%
  \BibitemOpen
  \bibfield  {author} {\bibinfo {author} {\bibfnamefont {S.}~\bibnamefont
  {Kuhn}}, \bibinfo {author} {\bibfnamefont {A.}~\bibnamefont {Kosloff}},
  \bibinfo {author} {\bibfnamefont {B.~A.}\ \bibnamefont {Stickler}}, \bibinfo
  {author} {\bibfnamefont {F.}~\bibnamefont {Patolsky}}, \bibinfo {author}
  {\bibfnamefont {K.}~\bibnamefont {Hornberger}}, \bibinfo {author}
  {\bibfnamefont {M.}~\bibnamefont {Arndt}},\ and\ \bibinfo {author}
  {\bibfnamefont {J.}~\bibnamefont {Millen}},\ }\href
  {https://doi.org/10.1364/optica.4.000356} {\bibfield  {journal} {\bibinfo
  {journal} {Optica}\ }\textbf {\bibinfo {volume} {4}},\ \bibinfo {pages} {356}
  (\bibinfo {year} {2017})}\BibitemShut {NoStop}%
\bibitem [{\citenamefont {Schäfer}\ \emph {et~al.}(2020)\citenamefont
  {Schäfer}, \citenamefont {Rudolph}, \citenamefont {Hornberger},\ and\
  \citenamefont {Stickler}}]{Schafer2020}%
  \BibitemOpen
  \bibfield  {author} {\bibinfo {author} {\bibfnamefont {J.}~\bibnamefont
  {Schäfer}}, \bibinfo {author} {\bibfnamefont {H.}~\bibnamefont {Rudolph}},
  \bibinfo {author} {\bibfnamefont {K.}~\bibnamefont {Hornberger}},\ and\
  \bibinfo {author} {\bibfnamefont {B.~A.}\ \bibnamefont {Stickler}},\
  }\href@noop {} {\bibinfo {title} {Cooling nanorotors by elliptic coherent
  scattering}} (\bibinfo {year} {2020}),\ \Eprint
  {https://arxiv.org/abs/arXiv:2006.04090} {arXiv:2006.04090} \BibitemShut
  {NoStop}%
\bibitem [{\citenamefont {Stickler}\ \emph {et~al.}(2021)\citenamefont
  {Stickler}, \citenamefont {Hornberger},\ and\ \citenamefont
  {Kim}}]{Stickler2021}%
  \BibitemOpen
  \bibfield  {author} {\bibinfo {author} {\bibfnamefont {B.~A.}\ \bibnamefont
  {Stickler}}, \bibinfo {author} {\bibfnamefont {K.}~\bibnamefont
  {Hornberger}},\ and\ \bibinfo {author} {\bibfnamefont {M.~S.}\ \bibnamefont
  {Kim}},\ }\href@noop {} {\bibinfo {title} {Quantum rotations of
  nanoparticles}} (\bibinfo {year} {2021}),\ \Eprint
  {https://arxiv.org/abs/arXiv:2102.00992} {arXiv:2102.00992} \BibitemShut
  {NoStop}%
\bibitem [{\citenamefont {Millen}\ \emph {et~al.}(2020)\citenamefont {Millen},
  \citenamefont {Monteiro}, \citenamefont {Pettit},\ and\ \citenamefont
  {Vamivakas}}]{Millen2020a}%
  \BibitemOpen
  \bibfield  {author} {\bibinfo {author} {\bibfnamefont {J.}~\bibnamefont
  {Millen}}, \bibinfo {author} {\bibfnamefont {T.~S.}\ \bibnamefont
  {Monteiro}}, \bibinfo {author} {\bibfnamefont {R.}~\bibnamefont {Pettit}},\
  and\ \bibinfo {author} {\bibfnamefont {A.~N.}\ \bibnamefont {Vamivakas}},\
  }\href {https://doi.org/10.1088/1361-6633/ab6100} {\bibfield  {journal}
  {\bibinfo  {journal} {Reports on Progress in Physics}\ }\textbf {\bibinfo
  {volume} {83}},\ \bibinfo {pages} {026401} (\bibinfo {year}
  {2020})}\BibitemShut {NoStop}%
\bibitem [{\citenamefont {Millen}\ and\ \citenamefont
  {Stickler}(2020)}]{Millen2020}%
  \BibitemOpen
  \bibfield  {author} {\bibinfo {author} {\bibfnamefont {J.}~\bibnamefont
  {Millen}}\ and\ \bibinfo {author} {\bibfnamefont {B.~A.}\ \bibnamefont
  {Stickler}},\ }\href {https://doi.org/10.1080/00107514.2020.1854497}
  {\bibfield  {journal} {\bibinfo  {journal} {Contemporary Physics}\ ,\
  \bibinfo {pages} {1}} (\bibinfo {year} {2020})}\BibitemShut {NoStop}%
\bibitem [{\citenamefont {Romero-Isart}\ \emph {et~al.}(2010)\citenamefont
  {Romero-Isart}, \citenamefont {Juan}, \citenamefont {Quidant},\ and\
  \citenamefont {Cirac}}]{RomeroIsart2010}%
  \BibitemOpen
  \bibfield  {author} {\bibinfo {author} {\bibfnamefont {O.}~\bibnamefont
  {Romero-Isart}}, \bibinfo {author} {\bibfnamefont {M.~L.}\ \bibnamefont
  {Juan}}, \bibinfo {author} {\bibfnamefont {R.}~\bibnamefont {Quidant}},\ and\
  \bibinfo {author} {\bibfnamefont {J.~I.}\ \bibnamefont {Cirac}},\ }\href
  {https://doi.org/10.1088/1367-2630/12/3/033015} {\bibfield  {journal}
  {\bibinfo  {journal} {New Journal of Physics}\ }\textbf {\bibinfo {volume}
  {12}},\ \bibinfo {pages} {033015} (\bibinfo {year} {2010})}\BibitemShut
  {NoStop}%
\bibitem [{\citenamefont {Romero-Isart}\ \emph
  {et~al.}(2011{\natexlab{a}})\citenamefont {Romero-Isart}, \citenamefont
  {Pflanzer}, \citenamefont {Blaser}, \citenamefont {Kaltenbaek}, \citenamefont
  {Kiesel}, \citenamefont {Aspelmeyer},\ and\ \citenamefont
  {Cirac}}]{RomeroIsart-superposition-2011}%
  \BibitemOpen
  \bibfield  {author} {\bibinfo {author} {\bibfnamefont {O.}~\bibnamefont
  {Romero-Isart}}, \bibinfo {author} {\bibfnamefont {A.~C.}\ \bibnamefont
  {Pflanzer}}, \bibinfo {author} {\bibfnamefont {F.}~\bibnamefont {Blaser}},
  \bibinfo {author} {\bibfnamefont {R.}~\bibnamefont {Kaltenbaek}}, \bibinfo
  {author} {\bibfnamefont {N.}~\bibnamefont {Kiesel}}, \bibinfo {author}
  {\bibfnamefont {M.}~\bibnamefont {Aspelmeyer}},\ and\ \bibinfo {author}
  {\bibfnamefont {J.~I.}\ \bibnamefont {Cirac}},\ }\href
  {https://doi.org/10.1103/PhysRevLett.107.020405} {\bibfield  {journal}
  {\bibinfo  {journal} {Phys. Rev. Lett.}\ }\textbf {\bibinfo {volume} {107}},\
  \bibinfo {pages} {020405} (\bibinfo {year} {2011}{\natexlab{a}})}\BibitemShut
  {NoStop}%
\bibitem [{\citenamefont {Weiss}\ \emph {et~al.}(2020)\citenamefont {Weiss},
  \citenamefont {Roda-Llordes}, \citenamefont {Torrontegui}, \citenamefont
  {Aspelmeyer},\ and\ \citenamefont {Romero-Isart}}]{Weiss2020}%
  \BibitemOpen
  \bibfield  {author} {\bibinfo {author} {\bibfnamefont {T.}~\bibnamefont
  {Weiss}}, \bibinfo {author} {\bibfnamefont {M.}~\bibnamefont {Roda-Llordes}},
  \bibinfo {author} {\bibfnamefont {E.}~\bibnamefont {Torrontegui}}, \bibinfo
  {author} {\bibfnamefont {M.}~\bibnamefont {Aspelmeyer}},\ and\ \bibinfo
  {author} {\bibfnamefont {O.}~\bibnamefont {Romero-Isart}},\ }\href@noop {}
  {\bibinfo {title} {Large quantum delocalization of a levitated nanoparticle
  using optimal control: Applications for force sensing and entangling via weak
  forces}} (\bibinfo {year} {2020}),\ \Eprint
  {https://arxiv.org/abs/arXiv:2012.12260} {arXiv:2012.12260} \BibitemShut
  {NoStop}%
\bibitem [{\citenamefont {Carlesso}\ and\ \citenamefont
  {Paternostro}(2020)}]{Carlesso2020}%
  \BibitemOpen
  \bibfield  {author} {\bibinfo {author} {\bibfnamefont {M.}~\bibnamefont
  {Carlesso}}\ and\ \bibinfo {author} {\bibfnamefont {M.}~\bibnamefont
  {Paternostro}},\ }in\ \href {https://doi.org/10.1007/978-3-030-46777-7_16}
  {\emph {\bibinfo {booktitle} {Fundamental Theories of Physics}}}\ (\bibinfo
  {publisher} {Springer International Publishing},\ \bibinfo {year} {2020})\
  pp.\ \bibinfo {pages} {205--215}\BibitemShut {NoStop}%
\bibitem [{\citenamefont {Ma}\ \emph {et~al.}(2020)\citenamefont {Ma},
  \citenamefont {Armata}, \citenamefont {Khosla},\ and\ \citenamefont
  {Kim}}]{Ma2020}%
  \BibitemOpen
  \bibfield  {author} {\bibinfo {author} {\bibfnamefont {Y.}~\bibnamefont
  {Ma}}, \bibinfo {author} {\bibfnamefont {F.}~\bibnamefont {Armata}}, \bibinfo
  {author} {\bibfnamefont {K.~E.}\ \bibnamefont {Khosla}},\ and\ \bibinfo
  {author} {\bibfnamefont {M.~S.}\ \bibnamefont {Kim}},\ }\href
  {https://doi.org/10.1103/PhysRevResearch.2.023208} {\bibfield  {journal}
  {\bibinfo  {journal} {Phys. Rev. Research}\ }\textbf {\bibinfo {volume}
  {2}},\ \bibinfo {pages} {023208} (\bibinfo {year} {2020})}\BibitemShut
  {NoStop}%
\bibitem [{\citenamefont {Brand\~ao}\ \emph {et~al.}(2020)\citenamefont
  {Brand\~ao}, \citenamefont {Suassuna}, \citenamefont {Melo},\ and\
  \citenamefont {Guerreiro}}]{Brandao2020}%
  \BibitemOpen
  \bibfield  {author} {\bibinfo {author} {\bibfnamefont {I.}~\bibnamefont
  {Brand\~ao}}, \bibinfo {author} {\bibfnamefont {B.}~\bibnamefont {Suassuna}},
  \bibinfo {author} {\bibfnamefont {B.}~\bibnamefont {Melo}},\ and\ \bibinfo
  {author} {\bibfnamefont {T.}~\bibnamefont {Guerreiro}},\ }\href
  {https://doi.org/10.1103/PhysRevResearch.2.043421} {\bibfield  {journal}
  {\bibinfo  {journal} {Phys. Rev. Research}\ }\textbf {\bibinfo {volume}
  {2}},\ \bibinfo {pages} {043421} (\bibinfo {year} {2020})}\BibitemShut
  {NoStop}%
\bibitem [{\citenamefont {Paternostro}\ \emph {et~al.}(2007)\citenamefont
  {Paternostro}, \citenamefont {Vitali}, \citenamefont {Gigan}, \citenamefont
  {Kim}, \citenamefont {Brukner}, \citenamefont {Eisert},\ and\ \citenamefont
  {Aspelmeyer}}]{Paternostro2007}%
  \BibitemOpen
  \bibfield  {author} {\bibinfo {author} {\bibfnamefont {M.}~\bibnamefont
  {Paternostro}}, \bibinfo {author} {\bibfnamefont {D.}~\bibnamefont {Vitali}},
  \bibinfo {author} {\bibfnamefont {S.}~\bibnamefont {Gigan}}, \bibinfo
  {author} {\bibfnamefont {M.~S.}\ \bibnamefont {Kim}}, \bibinfo {author}
  {\bibfnamefont {C.}~\bibnamefont {Brukner}}, \bibinfo {author} {\bibfnamefont
  {J.}~\bibnamefont {Eisert}},\ and\ \bibinfo {author} {\bibfnamefont
  {M.}~\bibnamefont {Aspelmeyer}},\ }\href
  {https://doi.org/10.1103/PhysRevLett.99.250401} {\bibfield  {journal}
  {\bibinfo  {journal} {Phys. Rev. Lett.}\ }\textbf {\bibinfo {volume} {99}},\
  \bibinfo {pages} {250401} (\bibinfo {year} {2007})}\BibitemShut {NoStop}%
\bibitem [{\citenamefont {Vitali}\ \emph {et~al.}(2007)\citenamefont {Vitali},
  \citenamefont {Gigan}, \citenamefont {Ferreira}, \citenamefont {B\"ohm},
  \citenamefont {Tombesi}, \citenamefont {Guerreiro}, \citenamefont {Vedral},
  \citenamefont {Zeilinger},\ and\ \citenamefont {Aspelmeyer}}]{Vitali2007}%
  \BibitemOpen
  \bibfield  {author} {\bibinfo {author} {\bibfnamefont {D.}~\bibnamefont
  {Vitali}}, \bibinfo {author} {\bibfnamefont {S.}~\bibnamefont {Gigan}},
  \bibinfo {author} {\bibfnamefont {A.}~\bibnamefont {Ferreira}}, \bibinfo
  {author} {\bibfnamefont {H.~R.}\ \bibnamefont {B\"ohm}}, \bibinfo {author}
  {\bibfnamefont {P.}~\bibnamefont {Tombesi}}, \bibinfo {author} {\bibfnamefont
  {A.}~\bibnamefont {Guerreiro}}, \bibinfo {author} {\bibfnamefont
  {V.}~\bibnamefont {Vedral}}, \bibinfo {author} {\bibfnamefont
  {A.}~\bibnamefont {Zeilinger}},\ and\ \bibinfo {author} {\bibfnamefont
  {M.}~\bibnamefont {Aspelmeyer}},\ }\href
  {https://doi.org/10.1103/PhysRevLett.98.030405} {\bibfield  {journal}
  {\bibinfo  {journal} {Phys. Rev. Lett.}\ }\textbf {\bibinfo {volume} {98}},\
  \bibinfo {pages} {030405} (\bibinfo {year} {2007})}\BibitemShut {NoStop}%
\bibitem [{\citenamefont {Krisnanda}\ \emph {et~al.}(2020)\citenamefont
  {Krisnanda}, \citenamefont {Tham}, \citenamefont {Paternostro},\ and\
  \citenamefont {Paterek}}]{Krisnanda2020}%
  \BibitemOpen
  \bibfield  {author} {\bibinfo {author} {\bibfnamefont {T.}~\bibnamefont
  {Krisnanda}}, \bibinfo {author} {\bibfnamefont {G.~Y.}\ \bibnamefont {Tham}},
  \bibinfo {author} {\bibfnamefont {M.}~\bibnamefont {Paternostro}},\ and\
  \bibinfo {author} {\bibfnamefont {T.}~\bibnamefont {Paterek}},\ }\bibfield
  {journal} {\bibinfo  {journal} {npj Quantum Information}\ }\textbf {\bibinfo
  {volume} {6}},\ \href {https://doi.org/10.1038/s41534-020-0243-y}
  {10.1038/s41534-020-0243-y} (\bibinfo {year} {2020})\BibitemShut {NoStop}%
\bibitem [{\citenamefont {Gut}\ \emph {et~al.}(2020)\citenamefont {Gut},
  \citenamefont {Winkler}, \citenamefont {Hoelscher-Obermaier}, \citenamefont
  {Hofer}, \citenamefont {Nia}, \citenamefont {Walk}, \citenamefont {Steffens},
  \citenamefont {Eisert}, \citenamefont {Wieczorek}, \citenamefont {Slater},
  \citenamefont {Aspelmeyer},\ and\ \citenamefont {Hammerer}}]{Gut2020}%
  \BibitemOpen
  \bibfield  {author} {\bibinfo {author} {\bibfnamefont {C.}~\bibnamefont
  {Gut}}, \bibinfo {author} {\bibfnamefont {K.}~\bibnamefont {Winkler}},
  \bibinfo {author} {\bibfnamefont {J.}~\bibnamefont {Hoelscher-Obermaier}},
  \bibinfo {author} {\bibfnamefont {S.~G.}\ \bibnamefont {Hofer}}, \bibinfo
  {author} {\bibfnamefont {R.~M.}\ \bibnamefont {Nia}}, \bibinfo {author}
  {\bibfnamefont {N.}~\bibnamefont {Walk}}, \bibinfo {author} {\bibfnamefont
  {A.}~\bibnamefont {Steffens}}, \bibinfo {author} {\bibfnamefont
  {J.}~\bibnamefont {Eisert}}, \bibinfo {author} {\bibfnamefont
  {W.}~\bibnamefont {Wieczorek}}, \bibinfo {author} {\bibfnamefont {J.~A.}\
  \bibnamefont {Slater}}, \bibinfo {author} {\bibfnamefont {M.}~\bibnamefont
  {Aspelmeyer}},\ and\ \bibinfo {author} {\bibfnamefont {K.}~\bibnamefont
  {Hammerer}},\ }\href {https://doi.org/10.1103/PhysRevResearch.2.033244}
  {\bibfield  {journal} {\bibinfo  {journal} {Phys. Rev. Research}\ }\textbf
  {\bibinfo {volume} {2}},\ \bibinfo {pages} {033244} (\bibinfo {year}
  {2020})}\BibitemShut {NoStop}%
\bibitem [{\citenamefont {Chauhan}\ \emph {et~al.}(2020)\citenamefont
  {Chauhan}, \citenamefont {{\v{C}}ernot{\'{\i}}k},\ and\ \citenamefont
  {Filip}}]{Chauhan2020}%
  \BibitemOpen
  \bibfield  {author} {\bibinfo {author} {\bibfnamefont {A.~K.}\ \bibnamefont
  {Chauhan}}, \bibinfo {author} {\bibfnamefont {O.}~\bibnamefont
  {{\v{C}}ernot{\'{\i}}k}},\ and\ \bibinfo {author} {\bibfnamefont
  {R.}~\bibnamefont {Filip}},\ }\href
  {https://doi.org/10.1088/1367-2630/abcce6} {\bibfield  {journal} {\bibinfo
  {journal} {New Journal of Physics}\ }\textbf {\bibinfo {volume} {22}},\
  \bibinfo {pages} {123021} (\bibinfo {year} {2020})}\BibitemShut {NoStop}%
\bibitem [{\citenamefont {Aggarwal}\ \emph {et~al.}(2020)\citenamefont
  {Aggarwal}, \citenamefont {Winstone}, \citenamefont {Teo}, \citenamefont
  {Baryakhtar}, \citenamefont {Larson}, \citenamefont {Kalogera},\ and\
  \citenamefont {Geraci}}]{Aggarwal2020searching}%
  \BibitemOpen
  \bibfield  {author} {\bibinfo {author} {\bibfnamefont {N.}~\bibnamefont
  {Aggarwal}}, \bibinfo {author} {\bibfnamefont {G.~P.}\ \bibnamefont
  {Winstone}}, \bibinfo {author} {\bibfnamefont {M.}~\bibnamefont {Teo}},
  \bibinfo {author} {\bibfnamefont {M.}~\bibnamefont {Baryakhtar}}, \bibinfo
  {author} {\bibfnamefont {S.~L.}\ \bibnamefont {Larson}}, \bibinfo {author}
  {\bibfnamefont {V.}~\bibnamefont {Kalogera}},\ and\ \bibinfo {author}
  {\bibfnamefont {A.~A.}\ \bibnamefont {Geraci}},\ }\href@noop {} {\bibinfo
  {title} {Searching for new physics with a levitated-sensor-based
  gravitational-wave detector}} (\bibinfo {year} {2020}),\ \Eprint
  {https://arxiv.org/abs/2010.13157} {arXiv:2010.13157 [gr-qc]} \BibitemShut
  {NoStop}%
\bibitem [{\citenamefont {Rakhubovsky}\ \emph {et~al.}(2020)\citenamefont
  {Rakhubovsky}, \citenamefont {Moore}, \citenamefont
  {Deli\ifmmode~\acute{c}\else \'{c}\fi{}}, \citenamefont {Kiesel},
  \citenamefont {Aspelmeyer},\ and\ \citenamefont {Filip}}]{Rakhubovsky2020}%
  \BibitemOpen
  \bibfield  {author} {\bibinfo {author} {\bibfnamefont {A.~A.}\ \bibnamefont
  {Rakhubovsky}}, \bibinfo {author} {\bibfnamefont {D.~W.}\ \bibnamefont
  {Moore}}, \bibinfo {author} {\bibfnamefont {U.~c.~v.}\ \bibnamefont
  {Deli\ifmmode~\acute{c}\else \'{c}\fi{}}}, \bibinfo {author} {\bibfnamefont
  {N.}~\bibnamefont {Kiesel}}, \bibinfo {author} {\bibfnamefont
  {M.}~\bibnamefont {Aspelmeyer}},\ and\ \bibinfo {author} {\bibfnamefont
  {R.}~\bibnamefont {Filip}},\ }\href
  {https://doi.org/10.1103/PhysRevApplied.14.054052} {\bibfield  {journal}
  {\bibinfo  {journal} {Phys. Rev. Applied}\ }\textbf {\bibinfo {volume}
  {14}},\ \bibinfo {pages} {054052} (\bibinfo {year} {2020})}\BibitemShut
  {NoStop}%
\bibitem [{\citenamefont {Windey}\ \emph {et~al.}(2019)\citenamefont {Windey},
  \citenamefont {Gonzalez-Ballestero}, \citenamefont {Maurer}, \citenamefont
  {Novotny}, \citenamefont {Romero-Isart},\ and\ \citenamefont
  {Reimann}}]{Windey2019}%
  \BibitemOpen
  \bibfield  {author} {\bibinfo {author} {\bibfnamefont {D.}~\bibnamefont
  {Windey}}, \bibinfo {author} {\bibfnamefont {C.}~\bibnamefont
  {Gonzalez-Ballestero}}, \bibinfo {author} {\bibfnamefont {P.}~\bibnamefont
  {Maurer}}, \bibinfo {author} {\bibfnamefont {L.}~\bibnamefont {Novotny}},
  \bibinfo {author} {\bibfnamefont {O.}~\bibnamefont {Romero-Isart}},\ and\
  \bibinfo {author} {\bibfnamefont {R.}~\bibnamefont {Reimann}},\ }\bibfield
  {journal} {\bibinfo  {journal} {Physical Review Letters}\ }\textbf {\bibinfo
  {volume} {122}},\ \href {https://doi.org/10.1103/physrevlett.122.123601}
  {10.1103/physrevlett.122.123601} (\bibinfo {year} {2019})\BibitemShut
  {NoStop}%
\bibitem [{\citenamefont {Deli{\'{c}}}\ \emph {et~al.}(2019)\citenamefont
  {Deli{\'{c}}}, \citenamefont {Reisenbauer}, \citenamefont {Grass},
  \citenamefont {Kiesel}, \citenamefont {Vuleti{\'{c}}},\ and\ \citenamefont
  {Aspelmeyer}}]{Delic2019}%
  \BibitemOpen
  \bibfield  {author} {\bibinfo {author} {\bibfnamefont {U.}~\bibnamefont
  {Deli{\'{c}}}}, \bibinfo {author} {\bibfnamefont {M.}~\bibnamefont
  {Reisenbauer}}, \bibinfo {author} {\bibfnamefont {D.}~\bibnamefont {Grass}},
  \bibinfo {author} {\bibfnamefont {N.}~\bibnamefont {Kiesel}}, \bibinfo
  {author} {\bibfnamefont {V.}~\bibnamefont {Vuleti{\'{c}}}},\ and\ \bibinfo
  {author} {\bibfnamefont {M.}~\bibnamefont {Aspelmeyer}},\ }\bibfield
  {journal} {\bibinfo  {journal} {Physical Review Letters}\ }\textbf {\bibinfo
  {volume} {122}},\ \href {https://doi.org/10.1103/physrevlett.122.123602}
  {10.1103/physrevlett.122.123602} (\bibinfo {year} {2019})\BibitemShut
  {NoStop}%
\bibitem [{\citenamefont {de~los R{\'{\i}}os~Sommer}\ \emph
  {et~al.}(2021)\citenamefont {de~los R{\'{\i}}os~Sommer}, \citenamefont
  {Meyer},\ and\ \citenamefont {Quidant}}]{DeLosRios2020}%
  \BibitemOpen
  \bibfield  {author} {\bibinfo {author} {\bibfnamefont {A.}~\bibnamefont
  {de~los R{\'{\i}}os~Sommer}}, \bibinfo {author} {\bibfnamefont
  {N.}~\bibnamefont {Meyer}},\ and\ \bibinfo {author} {\bibfnamefont
  {R.}~\bibnamefont {Quidant}},\ }\bibfield  {journal} {\bibinfo  {journal}
  {Nature Communications}\ }\textbf {\bibinfo {volume} {12}},\ \href
  {https://doi.org/10.1038/s41467-020-20419-2} {10.1038/s41467-020-20419-2}
  (\bibinfo {year} {2021})\BibitemShut {NoStop}%
\bibitem [{\citenamefont {Vuleti\ifmmode~\acute{c}\else \'{c}\fi{}}\ \emph
  {et~al.}(2001)\citenamefont {Vuleti\ifmmode~\acute{c}\else \'{c}\fi{}},
  \citenamefont {Chan},\ and\ \citenamefont {Black}}]{Vuletic2001}%
  \BibitemOpen
  \bibfield  {author} {\bibinfo {author} {\bibfnamefont {V.}~\bibnamefont
  {Vuleti\ifmmode~\acute{c}\else \'{c}\fi{}}}, \bibinfo {author} {\bibfnamefont
  {H.~W.}\ \bibnamefont {Chan}},\ and\ \bibinfo {author} {\bibfnamefont
  {A.~T.}\ \bibnamefont {Black}},\ }\href
  {https://doi.org/10.1103/PhysRevA.64.033405} {\bibfield  {journal} {\bibinfo
  {journal} {Phys. Rev. A}\ }\textbf {\bibinfo {volume} {64}},\ \bibinfo
  {pages} {033405} (\bibinfo {year} {2001})}\BibitemShut {NoStop}%
\bibitem [{\citenamefont {Toroš}\ \emph {et~al.}(2020)\citenamefont {Toroš},
  \citenamefont {Delić}, \citenamefont {Hales},\ and\ \citenamefont
  {Monteiro}}]{Toros2020}%
  \BibitemOpen
  \bibfield  {author} {\bibinfo {author} {\bibfnamefont {M.}~\bibnamefont
  {Toroš}}, \bibinfo {author} {\bibfnamefont {U.}~\bibnamefont {Delić}},
  \bibinfo {author} {\bibfnamefont {F.}~\bibnamefont {Hales}},\ and\ \bibinfo
  {author} {\bibfnamefont {T.~S.}\ \bibnamefont {Monteiro}},\ }\href@noop {}
  {\bibinfo {title} {Coherent scattering 2d cooling in levitated cavity
  optomechanics}} (\bibinfo {year} {2020}),\ \Eprint
  {https://arxiv.org/abs/arXiv:2012.15822} {arXiv:2012.15822} \BibitemShut
  {NoStop}%
\bibitem [{\citenamefont {Gonzalez-Ballestero}\ \emph
  {et~al.}(2019)\citenamefont {Gonzalez-Ballestero}, \citenamefont {Maurer},
  \citenamefont {Windey}, \citenamefont {Novotny}, \citenamefont {Reimann},\
  and\ \citenamefont {Romero-Isart}}]{Gonzalez-Ballestero2019}%
  \BibitemOpen
  \bibfield  {author} {\bibinfo {author} {\bibfnamefont {C.}~\bibnamefont
  {Gonzalez-Ballestero}}, \bibinfo {author} {\bibfnamefont {P.}~\bibnamefont
  {Maurer}}, \bibinfo {author} {\bibfnamefont {D.}~\bibnamefont {Windey}},
  \bibinfo {author} {\bibfnamefont {L.}~\bibnamefont {Novotny}}, \bibinfo
  {author} {\bibfnamefont {R.}~\bibnamefont {Reimann}},\ and\ \bibinfo {author}
  {\bibfnamefont {O.}~\bibnamefont {Romero-Isart}},\ }\bibfield  {journal}
  {\bibinfo  {journal} {Physical Review A}\ }\textbf {\bibinfo {volume}
  {100}},\ \href {https://doi.org/10.1103/physreva.100.013805}
  {10.1103/physreva.100.013805} (\bibinfo {year} {2019})\BibitemShut {NoStop}%
\bibitem [{\citenamefont {Cheng}\ \emph {et~al.}(2016)\citenamefont {Cheng},
  \citenamefont {Zhang}, \citenamefont {Zhou},\ and\ \citenamefont
  {Zhang}}]{Cheng2016}%
  \BibitemOpen
  \bibfield  {author} {\bibinfo {author} {\bibfnamefont {J.}~\bibnamefont
  {Cheng}}, \bibinfo {author} {\bibfnamefont {W.-Z.}\ \bibnamefont {Zhang}},
  \bibinfo {author} {\bibfnamefont {L.}~\bibnamefont {Zhou}},\ and\ \bibinfo
  {author} {\bibfnamefont {W.}~\bibnamefont {Zhang}},\ }\bibfield  {journal}
  {\bibinfo  {journal} {Scientific Reports}\ }\textbf {\bibinfo {volume} {6}},\
  \href {https://doi.org/10.1038/srep23678} {10.1038/srep23678} (\bibinfo
  {year} {2016})\BibitemShut {NoStop}%
\bibitem [{\citenamefont {Chen}\ \emph {et~al.}(2017)\citenamefont {Chen},
  \citenamefont {Lin}, \citenamefont {He},\ and\ \citenamefont
  {Lin}}]{Chen2017}%
  \BibitemOpen
  \bibfield  {author} {\bibinfo {author} {\bibfnamefont {Z.~X.}\ \bibnamefont
  {Chen}}, \bibinfo {author} {\bibfnamefont {Q.}~\bibnamefont {Lin}}, \bibinfo
  {author} {\bibfnamefont {B.}~\bibnamefont {He}},\ and\ \bibinfo {author}
  {\bibfnamefont {Z.~Y.}\ \bibnamefont {Lin}},\ }\href
  {https://doi.org/10.1364/oe.25.017237} {\bibfield  {journal} {\bibinfo
  {journal} {Optics Express}\ }\textbf {\bibinfo {volume} {25}},\ \bibinfo
  {pages} {17237} (\bibinfo {year} {2017})}\BibitemShut {NoStop}%
\bibitem [{\citenamefont {Li}\ \emph {et~al.}(2019)\citenamefont {Li},
  \citenamefont {Nie}, \citenamefont {Li},\ and\ \citenamefont
  {Chen}}]{Li2019}%
  \BibitemOpen
  \bibfield  {author} {\bibinfo {author} {\bibfnamefont {G.}~\bibnamefont
  {Li}}, \bibinfo {author} {\bibfnamefont {W.}~\bibnamefont {Nie}}, \bibinfo
  {author} {\bibfnamefont {X.}~\bibnamefont {Li}},\ and\ \bibinfo {author}
  {\bibfnamefont {A.}~\bibnamefont {Chen}},\ }\href
  {https://doi.org/10.1103/PhysRevA.100.063805} {\bibfield  {journal} {\bibinfo
   {journal} {Phys. Rev. A}\ }\textbf {\bibinfo {volume} {100}},\ \bibinfo
  {pages} {063805} (\bibinfo {year} {2019})}\BibitemShut {NoStop}%
\bibitem [{\citenamefont {Faroughi}\ \emph {et~al.}(2021)\citenamefont
  {Faroughi}, \citenamefont {Ahanj}, \citenamefont {Javidan},\ and\
  \citenamefont {Nazifkar}}]{Faroughi2021}%
  \BibitemOpen
  \bibfield  {author} {\bibinfo {author} {\bibfnamefont {Z.}~\bibnamefont
  {Faroughi}}, \bibinfo {author} {\bibfnamefont {A.}~\bibnamefont {Ahanj}},
  \bibinfo {author} {\bibfnamefont {K.}~\bibnamefont {Javidan}},\ and\ \bibinfo
  {author} {\bibfnamefont {S.}~\bibnamefont {Nazifkar}},\ }\href
  {https://doi.org/10.1007/s10773-020-04671-2} {\bibfield  {journal} {\bibinfo
  {journal} {International Journal of Theoretical Physics}\ }\textbf {\bibinfo
  {volume} {60}},\ \bibinfo {pages} {155} (\bibinfo {year} {2021})}\BibitemShut
  {NoStop}%
\bibitem [{\citenamefont {Rudolph}\ \emph {et~al.}(2020)\citenamefont
  {Rudolph}, \citenamefont {Hornberger},\ and\ \citenamefont
  {Stickler}}]{Hornberger2020}%
  \BibitemOpen
  \bibfield  {author} {\bibinfo {author} {\bibfnamefont {H.}~\bibnamefont
  {Rudolph}}, \bibinfo {author} {\bibfnamefont {K.}~\bibnamefont
  {Hornberger}},\ and\ \bibinfo {author} {\bibfnamefont {B.~A.}\ \bibnamefont
  {Stickler}},\ }\href {https://doi.org/10.1103/PhysRevA.101.011804} {\bibfield
   {journal} {\bibinfo  {journal} {Phys. Rev. A}\ }\textbf {\bibinfo {volume}
  {101}},\ \bibinfo {pages} {011804} (\bibinfo {year} {2020})}\BibitemShut
  {NoStop}%
\bibitem [{\citenamefont {Romero-Isart}\ \emph
  {et~al.}(2011{\natexlab{b}})\citenamefont {Romero-Isart}, \citenamefont
  {Pflanzer}, \citenamefont {Juan}, \citenamefont {Quidant}, \citenamefont
  {Kiesel}, \citenamefont {Aspelmeyer},\ and\ \citenamefont
  {Cirac}}]{Romero-Isart2011}%
  \BibitemOpen
  \bibfield  {author} {\bibinfo {author} {\bibfnamefont {O.}~\bibnamefont
  {Romero-Isart}}, \bibinfo {author} {\bibfnamefont {A.~C.}\ \bibnamefont
  {Pflanzer}}, \bibinfo {author} {\bibfnamefont {M.~L.}\ \bibnamefont {Juan}},
  \bibinfo {author} {\bibfnamefont {R.}~\bibnamefont {Quidant}}, \bibinfo
  {author} {\bibfnamefont {N.}~\bibnamefont {Kiesel}}, \bibinfo {author}
  {\bibfnamefont {M.}~\bibnamefont {Aspelmeyer}},\ and\ \bibinfo {author}
  {\bibfnamefont {J.~I.}\ \bibnamefont {Cirac}},\ }\bibfield  {journal}
  {\bibinfo  {journal} {Physical Review A}\ }\textbf {\bibinfo {volume} {83}},\
  \href {https://doi.org/10.1103/physreva.83.013803}
  {10.1103/physreva.83.013803} (\bibinfo {year}
  {2011}{\natexlab{b}})\BibitemShut {NoStop}%
\bibitem [{\citenamefont {Brandão}(2021)}]{numerical_toolbox}%
  \BibitemOpen
  \bibfield  {author} {\bibinfo {author} {\bibfnamefont {I.}~\bibnamefont
  {Brandão}},\ }\href@noop {} {\bibinfo {title} {Quantum gaussian information
  numerical toolbox}} (\bibinfo {year} {2021}),\ \bibinfo {note}
  {https://github.com/IgorBrandao42/Quantum-Gaussian-Information-Numerical-Toolbox.
  Retrieved 29/06/2021}\BibitemShut {NoStop}%
\bibitem [{\citenamefont {Jackson}(1998)}]{Jackson1998}%
  \BibitemOpen
  \bibfield  {author} {\bibinfo {author} {\bibfnamefont {J.}~\bibnamefont
  {Jackson}},\ }\href@noop {} {\emph {\bibinfo {title} {Classical
  Electrodynamics. New York: John Wiley\& Sons}}}\ (\bibinfo  {publisher}
  {Inc},\ \bibinfo {year} {1998})\BibitemShut {NoStop}%
\bibitem [{\citenamefont {Aspelmeyer}\ \emph {et~al.}(2014)\citenamefont
  {Aspelmeyer}, \citenamefont {Kippenberg},\ and\ \citenamefont
  {Marquardt}}]{Review_Aspelmeyer}%
  \BibitemOpen
  \bibfield  {author} {\bibinfo {author} {\bibfnamefont {M.}~\bibnamefont
  {Aspelmeyer}}, \bibinfo {author} {\bibfnamefont {T.~J.}\ \bibnamefont
  {Kippenberg}},\ and\ \bibinfo {author} {\bibfnamefont {F.}~\bibnamefont
  {Marquardt}},\ }\href {https://doi.org/10.1103/RevModPhys.86.1391} {\bibfield
   {journal} {\bibinfo  {journal} {Rev. Mod. Phys.}\ }\textbf {\bibinfo
  {volume} {86}},\ \bibinfo {pages} {1391} (\bibinfo {year}
  {2014})}\BibitemShut {NoStop}%
\bibitem [{\citenamefont {Chen}\ \emph {et~al.}(2020)\citenamefont {Chen},
  \citenamefont {Rossi}, \citenamefont {Mason},\ and\ \citenamefont
  {Schliesser}}]{Chen2020}%
  \BibitemOpen
  \bibfield  {author} {\bibinfo {author} {\bibfnamefont {J.}~\bibnamefont
  {Chen}}, \bibinfo {author} {\bibfnamefont {M.}~\bibnamefont {Rossi}},
  \bibinfo {author} {\bibfnamefont {D.}~\bibnamefont {Mason}},\ and\ \bibinfo
  {author} {\bibfnamefont {A.}~\bibnamefont {Schliesser}},\ }\bibfield
  {journal} {\bibinfo  {journal} {Nature Communications}\ }\textbf {\bibinfo
  {volume} {11}},\ \href {https://doi.org/10.1038/s41467-020-14768-1}
  {10.1038/s41467-020-14768-1} (\bibinfo {year} {2020})\BibitemShut {NoStop}%
\bibitem [{\citenamefont {Giovannetti}\ and\ \citenamefont
  {Vitali}(2001{\natexlab{a}})}]{Giovannetti2001}%
  \BibitemOpen
  \bibfield  {author} {\bibinfo {author} {\bibfnamefont {V.}~\bibnamefont
  {Giovannetti}}\ and\ \bibinfo {author} {\bibfnamefont {D.}~\bibnamefont
  {Vitali}},\ }\href {https://doi.org/10.1103/PhysRevA.63.023812} {\bibfield
  {journal} {\bibinfo  {journal} {Phys. Rev. A}\ }\textbf {\bibinfo {volume}
  {63}},\ \bibinfo {pages} {023812} (\bibinfo {year}
  {2001}{\natexlab{a}})}\BibitemShut {NoStop}%
\bibitem [{\citenamefont {Genes}\ \emph
  {et~al.}(2008{\natexlab{a}})\citenamefont {Genes}, \citenamefont {Mari},
  \citenamefont {Tombesi},\ and\ \citenamefont {Vitali}}]{Genes2008}%
  \BibitemOpen
  \bibfield  {author} {\bibinfo {author} {\bibfnamefont {C.}~\bibnamefont
  {Genes}}, \bibinfo {author} {\bibfnamefont {A.}~\bibnamefont {Mari}},
  \bibinfo {author} {\bibfnamefont {P.}~\bibnamefont {Tombesi}},\ and\ \bibinfo
  {author} {\bibfnamefont {D.}~\bibnamefont {Vitali}},\ }\href
  {https://doi.org/10.1103/PhysRevA.78.032316} {\bibfield  {journal} {\bibinfo
  {journal} {Phys. Rev. A}\ }\textbf {\bibinfo {volume} {78}},\ \bibinfo
  {pages} {032316} (\bibinfo {year} {2008}{\natexlab{a}})}\BibitemShut
  {NoStop}%
\bibitem [{\citenamefont {Genes}\ \emph
  {et~al.}(2008{\natexlab{b}})\citenamefont {Genes}, \citenamefont {Vitali},
  \citenamefont {Tombesi}, \citenamefont {Gigan},\ and\ \citenamefont
  {Aspelmeyer}}]{Genes2009}%
  \BibitemOpen
  \bibfield  {author} {\bibinfo {author} {\bibfnamefont {C.}~\bibnamefont
  {Genes}}, \bibinfo {author} {\bibfnamefont {D.}~\bibnamefont {Vitali}},
  \bibinfo {author} {\bibfnamefont {P.}~\bibnamefont {Tombesi}}, \bibinfo
  {author} {\bibfnamefont {S.}~\bibnamefont {Gigan}},\ and\ \bibinfo {author}
  {\bibfnamefont {M.}~\bibnamefont {Aspelmeyer}},\ }\href
  {https://doi.org/10.1103/PhysRevA.77.033804} {\bibfield  {journal} {\bibinfo
  {journal} {Phys. Rev. A}\ }\textbf {\bibinfo {volume} {77}},\ \bibinfo
  {pages} {033804} (\bibinfo {year} {2008}{\natexlab{b}})}\BibitemShut
  {NoStop}%
\bibitem [{\citenamefont {Sommer}\ and\ \citenamefont
  {Genes}(2019)}]{Sommer2019}%
  \BibitemOpen
  \bibfield  {author} {\bibinfo {author} {\bibfnamefont {C.}~\bibnamefont
  {Sommer}}\ and\ \bibinfo {author} {\bibfnamefont {C.}~\bibnamefont {Genes}},\
  }\href {https://doi.org/10.1103/PhysRevLett.123.203605} {\bibfield  {journal}
  {\bibinfo  {journal} {Phys. Rev. Lett.}\ }\textbf {\bibinfo {volume} {123}},\
  \bibinfo {pages} {203605} (\bibinfo {year} {2019})}\BibitemShut {NoStop}%
\bibitem [{\citenamefont {Gardiner}\ and\ \citenamefont
  {Collett}(1985)}]{Gardiner1985}%
  \BibitemOpen
  \bibfield  {author} {\bibinfo {author} {\bibfnamefont {C.~W.}\ \bibnamefont
  {Gardiner}}\ and\ \bibinfo {author} {\bibfnamefont {M.~J.}\ \bibnamefont
  {Collett}},\ }\href {https://doi.org/10.1103/physreva.31.3761} {\bibfield
  {journal} {\bibinfo  {journal} {Physical Review A}\ }\textbf {\bibinfo
  {volume} {31}},\ \bibinfo {pages} {3761} (\bibinfo {year}
  {1985})}\BibitemShut {NoStop}%
\bibitem [{\citenamefont {Giovannetti}\ and\ \citenamefont
  {Vitali}(2001{\natexlab{b}})}]{Vitali2001}%
  \BibitemOpen
  \bibfield  {author} {\bibinfo {author} {\bibfnamefont {V.}~\bibnamefont
  {Giovannetti}}\ and\ \bibinfo {author} {\bibfnamefont {D.}~\bibnamefont
  {Vitali}},\ }\href {https://doi.org/10.1103/PhysRevA.63.023812} {\bibfield
  {journal} {\bibinfo  {journal} {Phys. Rev. A}\ }\textbf {\bibinfo {volume}
  {63}},\ \bibinfo {pages} {023812} (\bibinfo {year}
  {2001}{\natexlab{b}})}\BibitemShut {NoStop}%
\bibitem [{\citenamefont {Weedbrook}\ \emph {et~al.}(2012)\citenamefont
  {Weedbrook}, \citenamefont {Pirandola}, \citenamefont {Garc\'{\i}a-Patr\'on},
  \citenamefont {Cerf}, \citenamefont {Ralph}, \citenamefont {Shapiro},\ and\
  \citenamefont {Lloyd}}]{Lloyd2012}%
  \BibitemOpen
  \bibfield  {author} {\bibinfo {author} {\bibfnamefont {C.}~\bibnamefont
  {Weedbrook}}, \bibinfo {author} {\bibfnamefont {S.}~\bibnamefont
  {Pirandola}}, \bibinfo {author} {\bibfnamefont {R.}~\bibnamefont
  {Garc\'{\i}a-Patr\'on}}, \bibinfo {author} {\bibfnamefont {N.~J.}\
  \bibnamefont {Cerf}}, \bibinfo {author} {\bibfnamefont {T.~C.}\ \bibnamefont
  {Ralph}}, \bibinfo {author} {\bibfnamefont {J.~H.}\ \bibnamefont {Shapiro}},\
  and\ \bibinfo {author} {\bibfnamefont {S.}~\bibnamefont {Lloyd}},\ }\href
  {https://doi.org/10.1103/RevModPhys.84.621} {\bibfield  {journal} {\bibinfo
  {journal} {Rev. Mod. Phys.}\ }\textbf {\bibinfo {volume} {84}},\ \bibinfo
  {pages} {621} (\bibinfo {year} {2012})}\BibitemShut {NoStop}%
\bibitem [{\citenamefont {Groisman}\ \emph {et~al.}(2005)\citenamefont
  {Groisman}, \citenamefont {Popescu},\ and\ \citenamefont
  {Winter}}]{Groisman2005}%
  \BibitemOpen
  \bibfield  {author} {\bibinfo {author} {\bibfnamefont {B.}~\bibnamefont
  {Groisman}}, \bibinfo {author} {\bibfnamefont {S.}~\bibnamefont {Popescu}},\
  and\ \bibinfo {author} {\bibfnamefont {A.}~\bibnamefont {Winter}},\ }\href
  {https://doi.org/10.1103/PhysRevA.72.032317} {\bibfield  {journal} {\bibinfo
  {journal} {Phys. Rev. A}\ }\textbf {\bibinfo {volume} {72}},\ \bibinfo
  {pages} {032317} (\bibinfo {year} {2005})}\BibitemShut {NoStop}%
\bibitem [{\citenamefont {\ifmmode~\check{C}\else \v{C}\fi{}ernot\'{\i}k}\ and\
  \citenamefont {Filip}(2020)}]{Radim2020}%
  \BibitemOpen
  \bibfield  {author} {\bibinfo {author} {\bibfnamefont {O.~c.~v.}\
  \bibnamefont {\ifmmode~\check{C}\else \v{C}\fi{}ernot\'{\i}k}}\ and\ \bibinfo
  {author} {\bibfnamefont {R.}~\bibnamefont {Filip}},\ }\href
  {https://doi.org/10.1103/PhysRevResearch.2.013052} {\bibfield  {journal}
  {\bibinfo  {journal} {Phys. Rev. Research}\ }\textbf {\bibinfo {volume}
  {2}},\ \bibinfo {pages} {013052} (\bibinfo {year} {2020})}\BibitemShut
  {NoStop}%
\bibitem [{\citenamefont {Pflanzer}\ \emph {et~al.}(2012)\citenamefont
  {Pflanzer}, \citenamefont {Romero-Isart},\ and\ \citenamefont
  {Cirac}}]{PFLANZER2012}%
  \BibitemOpen
  \bibfield  {author} {\bibinfo {author} {\bibfnamefont {A.~C.}\ \bibnamefont
  {Pflanzer}}, \bibinfo {author} {\bibfnamefont {O.}~\bibnamefont
  {Romero-Isart}},\ and\ \bibinfo {author} {\bibfnamefont {J.~I.}\ \bibnamefont
  {Cirac}},\ }\href {https://doi.org/10.1103/PhysRevA.86.013802} {\bibfield
  {journal} {\bibinfo  {journal} {Phys. Rev. A}\ }\textbf {\bibinfo {volume}
  {86}},\ \bibinfo {pages} {013802} (\bibinfo {year} {2012})}\BibitemShut
  {NoStop}%
\bibitem [{\citenamefont {Deli{\'{c}}}\ \emph
  {et~al.}(2020{\natexlab{b}})\citenamefont {Deli{\'{c}}}, \citenamefont
  {Grass}, \citenamefont {Reisenbauer}, \citenamefont {Damm}, \citenamefont
  {Weitz}, \citenamefont {Kiesel},\ and\ \citenamefont
  {Aspelmeyer}}]{Delic2020b}%
  \BibitemOpen
  \bibfield  {author} {\bibinfo {author} {\bibfnamefont {U.}~\bibnamefont
  {Deli{\'{c}}}}, \bibinfo {author} {\bibfnamefont {D.}~\bibnamefont {Grass}},
  \bibinfo {author} {\bibfnamefont {M.}~\bibnamefont {Reisenbauer}}, \bibinfo
  {author} {\bibfnamefont {T.}~\bibnamefont {Damm}}, \bibinfo {author}
  {\bibfnamefont {M.}~\bibnamefont {Weitz}}, \bibinfo {author} {\bibfnamefont
  {N.}~\bibnamefont {Kiesel}},\ and\ \bibinfo {author} {\bibfnamefont
  {M.}~\bibnamefont {Aspelmeyer}},\ }\href
  {https://doi.org/10.1088/2058-9565/ab7989} {\bibfield  {journal} {\bibinfo
  {journal} {Quantum Science and Technology}\ }\textbf {\bibinfo {volume}
  {5}},\ \bibinfo {pages} {025006} (\bibinfo {year}
  {2020}{\natexlab{b}})}\BibitemShut {NoStop}%
\bibitem [{\citenamefont {Adesso}\ and\ \citenamefont
  {Illuminati}(2007)}]{Adesso2007}%
  \BibitemOpen
  \bibfield  {author} {\bibinfo {author} {\bibfnamefont {G.}~\bibnamefont
  {Adesso}}\ and\ \bibinfo {author} {\bibfnamefont {F.}~\bibnamefont
  {Illuminati}},\ }\href {https://doi.org/10.1088/1751-8113/40/28/s01}
  {\bibfield  {journal} {\bibinfo  {journal} {Journal of Physics A:
  Mathematical and Theoretical}\ }\textbf {\bibinfo {volume} {40}},\ \bibinfo
  {pages} {7821} (\bibinfo {year} {2007})}\BibitemShut {NoStop}%
\bibitem [{\citenamefont {Herbut}(2004)}]{Herbut2004}%
  \BibitemOpen
  \bibfield  {author} {\bibinfo {author} {\bibfnamefont {F.}~\bibnamefont
  {Herbut}},\ }\href {https://doi.org/10.1088/0305-4470/37/10/016} {\bibfield
  {journal} {\bibinfo  {journal} {Journal of Physics A: Mathematical and
  General}\ }\textbf {\bibinfo {volume} {37}},\ \bibinfo {pages} {3535}
  (\bibinfo {year} {2004})}\BibitemShut {NoStop}%
\bibitem [{\citenamefont {Jain}\ \emph {et~al.}(2016)\citenamefont {Jain},
  \citenamefont {Gieseler}, \citenamefont {Moritz}, \citenamefont {Dellago},
  \citenamefont {Quidant},\ and\ \citenamefont {Novotny}}]{Jain2016}%
  \BibitemOpen
  \bibfield  {author} {\bibinfo {author} {\bibfnamefont {V.}~\bibnamefont
  {Jain}}, \bibinfo {author} {\bibfnamefont {J.}~\bibnamefont {Gieseler}},
  \bibinfo {author} {\bibfnamefont {C.}~\bibnamefont {Moritz}}, \bibinfo
  {author} {\bibfnamefont {C.}~\bibnamefont {Dellago}}, \bibinfo {author}
  {\bibfnamefont {R.}~\bibnamefont {Quidant}},\ and\ \bibinfo {author}
  {\bibfnamefont {L.}~\bibnamefont {Novotny}},\ }\href
  {https://doi.org/10.1103/PhysRevLett.116.243601} {\bibfield  {journal}
  {\bibinfo  {journal} {Phys. Rev. Lett.}\ }\textbf {\bibinfo {volume} {116}},\
  \bibinfo {pages} {243601} (\bibinfo {year} {2016})}\BibitemShut {NoStop}%
\bibitem [{\citenamefont {Tebbenjohanns}\ \emph {et~al.}(2021)\citenamefont
  {Tebbenjohanns}, \citenamefont {Mattana}, \citenamefont {Rossi},
  \citenamefont {Frimmer},\ and\ \citenamefont {Novotny}}]{Tebbenjohanns2021}%
  \BibitemOpen
  \bibfield  {author} {\bibinfo {author} {\bibfnamefont {F.}~\bibnamefont
  {Tebbenjohanns}}, \bibinfo {author} {\bibfnamefont {M.~L.}\ \bibnamefont
  {Mattana}}, \bibinfo {author} {\bibfnamefont {M.}~\bibnamefont {Rossi}},
  \bibinfo {author} {\bibfnamefont {M.}~\bibnamefont {Frimmer}},\ and\ \bibinfo
  {author} {\bibfnamefont {L.}~\bibnamefont {Novotny}},\ }\href@noop {}
  {\bibinfo {title} {Quantum control of a nanoparticle optically levitated in
  cryogenic free space}} (\bibinfo {year} {2021}),\ \Eprint
  {https://arxiv.org/abs/2103.03853} {arXiv:2103.03853 [quant-ph]} \BibitemShut
  {NoStop}%
\bibitem [{\citenamefont {Tebbenjohanns}\ \emph {et~al.}(2020)\citenamefont
  {Tebbenjohanns}, \citenamefont {Frimmer}, \citenamefont {Jain}, \citenamefont
  {Windey},\ and\ \citenamefont {Novotny}}]{Tebbenjohanns2020}%
  \BibitemOpen
  \bibfield  {author} {\bibinfo {author} {\bibfnamefont {F.}~\bibnamefont
  {Tebbenjohanns}}, \bibinfo {author} {\bibfnamefont {M.}~\bibnamefont
  {Frimmer}}, \bibinfo {author} {\bibfnamefont {V.}~\bibnamefont {Jain}},
  \bibinfo {author} {\bibfnamefont {D.}~\bibnamefont {Windey}},\ and\ \bibinfo
  {author} {\bibfnamefont {L.}~\bibnamefont {Novotny}},\ }\href
  {https://doi.org/10.1103/PhysRevLett.124.013603} {\bibfield  {journal}
  {\bibinfo  {journal} {Phys. Rev. Lett.}\ }\textbf {\bibinfo {volume} {124}},\
  \bibinfo {pages} {013603} (\bibinfo {year} {2020})}\BibitemShut {NoStop}%
\end{thebibliography}%

\newpage
\onecolumngrid
\appendix

\section{Hamiltonian derivation} \label{sec:appendix_Hamiltonian_details}

The free energy for the NPs and the the EM field \cite{Jackson1998} are, respectively,  given by

\begin{align}
    \H_{\mathrm{NP}} &= \sum_{j=1}^N  \frac{\PP_{j}^2}{2m_j}  \, , \\
    \H_{\rm field} &= \frac{\epsilon_0}{2} \int \hat{E}^2(\bm{r}) + c^2 \hat{B}^2(\bm{r}) d^3 \bm{r} \simeq \hbar\omega\cav\a\dager\a\; , \label{eq:field_hamiltonian}
\end{align}

\noindent where, in the second equation, we made use of the approximation of the EM field given in Eq. \eqref{eq:total_em_field} and neglected constant energies contributions, as will be done hereforth. The interaction Hamiltonian, Equation \eqref{eq:Int_Hamiltonian}, can be broken down into three pieces:



\begin{equation}
    \H_{\mathrm{int}} = \sum_{j=1}^{N} -\frac{\alpha_j}{2} \left \{ \vert\vec{E}\cav(\rr_j)\vert^2 + 2\text{Re}\Big(\vec{E}\cav(\rr_j)\vec{E}^*\twj(\rr_j)\Big)
      + \vert\vec{E}\twi(\rr_j)\vert^2 \right \} \, .
    %
\end{equation}


In order to evaluate the final form of the interaction energy, we use the definitions given to the cavity's and OTs' fields given by Eqs.  \eqref{eq:tweezer_electric_field} and \eqref{eq:cavity_electric_field}, respectively. The gaussian tweezers will effectively trap the NPs close to their focus, confining them close to $\bm{R}_{0,j}$, thus we can approximate the interaction Hamiltonian through a series expansion around $(\hat{X}_j, \hat{Y}_j, \hat{Z}_j) = \bm{0}$ in each term. Moreover, as the optical modes' frequencies ($\omega\twj \approx \omega\cav$) are much higher than the coupling rates present in the system , we also take a rotating-wave approximation (RWA) at these frequencies in each interaction term \cite{Gonzalez-Ballestero2019}. Finally, we disregard the constant energy shifts as they do not affect the dynamics of the system.

The tweezer-tweezer interaction terms yields a $3D$ trapping potential



\begin{equation}
    -\frac{\alpha}{2} \vert\hat{E}\twj(\RR)\vert^2 \approx \sum_{\substack{i = \\ x,y,z}} \frac{m \Omega_{j,i}^2}{2}\RR^2_{j,i}
\end{equation}



The cavity-cavity interaction results in three terms:

\begin{equation}
    -\half\alpha_j\vert\bm{E}_{\rm cav}(\rr_j)\vert^2 
    \approx -\hbar\delta_j \cos^2(k\cav x_{0,j})\, \a\dager\a 
    +\hbar k\cav\delta_j\sin(2k\cav x_{0,j})\hat{X}_j\bigg[ \a\dager\a + \half \bigg] \, ,
\end{equation}

\noindent one proportional only to the cavity's number operator, responsible for the cavity frequency shift due to the presence of the $j$-th particle, other proportional to the $j$-th NP position quadrature, acting as a constant drive in the NP's momentum, and an interacting term resulting in the radiation pressure effect on the $j$-th NP by the cavity field. 

Finally, the cavity-tweezers interaction gives rise to the coherent scattering interaction \cite{Gonzalez-Ballestero2019}, effectively 2D coupling the NPs with the cavity field and a drive in the cavity field. The nature of this interaction is due to the $j$-th tweezer's photons being coherently scattered by the particle inside the cavity, populating it \cite{Vuletic2001, Gonzalez-Ballestero2019, Delic2019}:

\begin{align}
    -\alpha_j\text{Re}\Big(\bm{\hat{E}}\cav(\rr_j)\bm{E}^*\twj(\rr_j)\Big) \approx &-\alpha_j\epsilon\tw \epsilon\cav\left( \a\dager + \a \right)
   \Bigg\{ \cos(\omega\tw t)  \bigg( \cos(k\cav x_{0,j}) - k\cav \hat{X}_j\sin(k\cav x_{0,j})\bigg) \nonumber\\
   &-\hat{Z}_j\frac{k\tw z_{R_j} - 1}{z_{R_j}}\sin(\omega\tw t)\cos(k\cav x_{0,j})  \Bigg\} \sin(\theta_j) \, .
\end{align}

\noindent Note that this last term has a time dependency, which we can get rid off by moving into a rotating reference frame at the frequency of the tweezers according to

\begin{equation*}
    \H \ra \U \H \U\dager - i\hbar\U\frac{\partial \U\dager}{\partial t} \, ,
\end{equation*}

\noindent where $\U(t) \equiv \exp\big(i\omega\tw \a\dager\a t\big)$. We apply another RWA by neglecting all rapidly oscillating terms at frequency $2\omega\tw$.

Gathering all these terms results in the general Hamiltonian for this system, presented in Equation \eqref{eq:Full_Hamiltonian}.


\section{Interaction with free field} \label{appendix:free_field}

The free EM modes can be described by

\begin{align}
    \bm{\hat{E}}_{\rm free}(\r) &= \sum_{\bm{k},\bm{e}}\epsilon_{\bm{k}}(\bm{e}_{\bm{k}}\me^{i\bm{k}\cdot\bm{r}}\cc+H.c.) = \sum_l\epsilon_l(\bm{e}_l \me^{i\bm{k}\cdot\bm{r}}\cl+H.c.) \, , \label{eq:free_electric_field}
\end{align}
\noindent where $\epsilon_l = \sqrt{\frac{\hbar\omega_l}{2\epsilon_0V_{\rm f}}}$,  $V_{\rm f}$ stand for the quantization volume and $\cc$ is the annihilation operator of a free EM mode with wave-vector $\bm{k}$ and polarization $\bm{\epsilon}_{\bm{k}}$. To simplify the notation, the index $l$ is used to denote the set $\{\bm{k},\bm{e}_{\bm{k}}\}$.

If we were to consider this additional EM field, the field Hamiltonian, in Equation \eqref{eq:field_hamiltonian}, and the interaction Hamiltonian, in Equation \eqref{eq:Int_Hamiltonian}, would transform into
\begin{align}
    \H_{\mathrm{field}}'  = \H_{\mathrm{field}} &+ \hbar \sum_l \omega_l \cl \dager\cl \, , \\
    \H_{\mathrm{int},j}'  = \H_{\mathrm{int},j} &-\half\alpha_j\Bigg(\vert\bm{E}\free(\rr_j)\vert^2 +2 Re\left\{\bm{E}\free(\rr_j)\bm{E}^*\twj(\rr_j)\right\} + 2 Re\left\{\bm{E}\free(\rr_j)\bm{E}^*\cav(\rr_j)\right\}\Bigg)  \, . \label{eq:interaction_full}
\end{align}
Consequently, the extra interaction terms that would appear on the complete Hamiltonian are
\begin{align}
    \H_{\mathrm{f-f},j}^{\rm int} &= -\half\alpha_j\vert\bm{E}_{\rm free}(\rr_j)\vert^2 = -\frac{\alpha_j\hbar}{4\epsilon_0V_{\rm f}}\sum_{l} \sum_{l'}
    \sqrt{\omega_{l}\omega_{l'}}\bm{e}_{l}\cdot\bm{e}_{l'} (\me^{i\bm{k}\cdot\rr_j}\cl+H.c.)
    (\me^{i\bm{k}'\cdot\rr_j}\hat{c}_{l'}+H.c.) \, ,
\end{align}
\noindent It was shown that in the long-wavelength approximation $\H_{\rm f-f,j}^{\rm int}$ can be safely neglected \cite{PFLANZER2012}. Moreover, the cavity-free fields interaction is
\begin{align}
    \hat{H}_{\mathrm{c-f},j}^{\rm int} &\approx -\hbar\sum_{l}G_{\rm cf}(l)\left( \a\dager + \a \right)(\me^{i\bm{k}\cdot\bm{R}_{0,j}}\cl+\me^{-i\bm{k}\cdot\bm{R}_{0,j}}\cl\dager)\cos(k\cav x_{0,j}) \, ,
\end{align}
\noindent where $G_{\rm cf}(l) = \alpha_j\epsilon_{l}\epsilon\cav\bm{e}_{l}\cdot\bm{e}_{y}/\hbar$. Notice that this terms dies out when we place the mean position of NP at the cavity nodes. Lastly, the tweezers-free fields interaction is the source of recoil heating \cite{DeLosRios2020} of the NP as they incoherently scatters light off the tweezers into free space. These interactions are given by
\begin{align}
    \hat{H}_{\mathrm{t-f},j}^{\rm int} \approx -\hbar\sum_{l}G_{\rm tf}(l)\Bigg[&
    (\me^{i\bm{k}\cdot\bm{R}_{0,j}}\cl+\me^{-i\bm{k}\cdot\bm{R}_{0,j}}\cl\dager) \left(\cos(\omega\tw t) - \z_j\frac{k\tw z_{\mathrm{R},j} - 1}{z_{\mathrm{R},j}}\sin(\omega\tw t) \right) \nonumber\\
        & +i\bm{k}\cdot\RR_j(\me^{i\bm{k}\cdot\bm{R}_{0,j}}\cl-\me^{-i\bm{k}\cdot\bm{R}_{0,j}}\cl\dager)\cos(\omega\tw t)
    \Bigg] \, ,
\end{align}
\noindent where $G_{\rm tf}(l) = \alpha_j\epsilon_{l}\epsilon\tw \bm{e}_{l}\cdot\bm{e}_y/\hbar$. By moving this term into the rotating reference frame at the tweezers' frequency $\omega\tw$, through the operator $\U\free(t) \equiv \exp\big(i\omega\tw t \sum_{l}\cl\dager\cl \big)$, we would also introduce a shifted free-field frequency $\Delta_l = \omega_l-\omega\tw$ and the field Hamiltonian and effective interaction with the free field would become

\begin{align}
 &\H_{\mathrm{field}}'  \ra \H_{\mathrm{field}}' + \hbar \sum_l \Delta_l \cl \dager\cl \, , \\
    &\hat{H}_{\mathrm{t-f},j}^{\rm int} \rightarrow -\hbar\sum_{l}G_{\rm tf}(l)\Bigg[
    \bigg(\cl \me^{i\bm{k}\cdot\bm{R}_{0,j}}+\cl\dager \me^{-i\bm{k}\cdot\bm{R}_{0,j}}\bigg) -i\bigg( \cl\dager \me^{-i\bm{k}\cdot\bm{R}_{0,j}} - \cl \me^{i\bm{k}\cdot\bm{R}_{0,j}} \bigg) \bigg(\bm{k}\cdot\RR_j + \z_j\frac{k\tw z_{\mathrm{R},j} - 1}{z_{\mathrm{R},j}}\bigg)
    \Bigg] \, .\nonumber 
\end{align}
We observe a 3D coupling between the nanoparticles with the free electric field. For typical experimental values \cite{Delic2020a, Delic2020b}, one usually has $w_{0,j} \approx \SI{0.7}{\mu\meter}$ such that $\frac{k\tw z_{\mathrm{R},j} - 1}{z_{\mathrm{R},j}} \approx k\tw \simeq 5.9 \cdot \SI[parse-numbers=false]{10^{-3}}{\nm^{-1}}$, $\omega_j \approx 2\pi \cdot\SI{305}{\kHz}$, and the mass of a single nanoparticle $m_j \approx \SI{2.83}{\femto\gram}$ such that $x_{\mathrm{ZPF}, j} \approx \SI{3.1}{\pico\metre}$. Therefore, for a particle in the motional ground state, it is expected that its coupling to the free field plays a negligible role compared to the first term in the Equation above, as $k\tw \cdot x_{\mathrm{ZPF}, j} \sim k \cdot x_{\mathrm{ZPF}, j} \ll 1$. 

If we consider a more energetic initial quantum state for the nanoparticles, say a thermal state at temperature $T = \SI[parse-numbers=false]{10}{K}$, through the equipartition theorem we could assert that initially $\sqrt{ \langle x_j^2 \rangle} = \sqrt{k_B T/(m_j\omega_j^2)} \simeq \SI[parse-numbers=false]{5.85}{\nm}$, such that we would still have $k\tw \cdot \sqrt{ \langle x_j^2 \rangle}  \sim k \cdot \sqrt{ \langle x_j^2 \rangle} \ll 1$.

Therefore, we can safely ignore the nanoparticles' interaction with the free field for sufficiently cooled down particles, making the free field  evolution uncoupled from the nanoparticles. Thus we may ignore it in the dynamics of the system as studied in the main text. Note that however small this interaction, we still consider its decoherence effects on the NPs in Appendix \ref{appendix:decoherence}. 

\section{Gaussian Quantum Information Toolbox} \label{appendix:gaussian_toolbox}

A quantum system whose states live in a infinite-dimensional Hilbert space described by observables with continuous spectra is called a \textit{continuous variable systems} \cite{Lloyd2012}.
The physical setup under study falls into such category: it is comprised of $M=N+1$ modes, one optical field and $N$ mechanical modes, each with their corresponding annihilation and creation operator. These can be conventionally arranged in a vectorial operator $\bm{\hat{c}} = (\a, \a\dager, \b_1, \b_1\dager, \b_2, \b_2\dager, \ldots)^T$ which satisfy the bosonic commutation relations expressed as $ [\hat{c}_j,\hat{c}_k] = \Omega_{jk} $ and it immediately follows that the vectorial operator $\bm{\hat{X}}$, defined in Section \ref{sec:open_dynamics}, satisfy the commutation relations $[\hat{X}_j,\hat{X}_k] = 2i\,\Omega_{jk}$,
where $j,k = 1, \ldots, 2M$ and $\Omega$ is the symplectic form $2M\times2M$ matrix given by 
\begin{equation}
    \Omega = \bigotimes_{k=1}^M \begin{bmatrix}
    0 & 1 \\
    -1 & 0
    \end{bmatrix} \, .
\end{equation}

\subsection*{Gaussian states}

We are interested in a special class of states for such system called \textit{gaussian states}, whose Wigner function is Gaussian; here we refer the more interested reader once again to \cite{Lloyd2012} for a more complete description of this topic. The crucial property of these states for us is the fact that a gaussian state with density matrix $\rho$ can be completely characterized by its first, $\langle \bm{\hat{X}} \rangle = \mathrm{tr}(\bm{\hat{X}}\rho)$, and second moments represented by its covariance matrix, $V_{i,j} = \frac{1}{2}\langle \hat{X}_i \hat{X}_j + \hat{X}_j \hat{X}_i\rangle$. This greatly simplifies our numerical treatment of these system, because instead of dealing with an infinite-dimensional density matrix, we need only to worry about a $2M\times2M$ matrix. In the following, we present the theoretical tools for gaussian states used in our simulations.

\subsection*{Partial trace}

Consider a density matrix $\rho_{AB}$  describing a multipartite gaussian state, which we choose to subdivide into the subsystems $A$ and $B$, respectively, with $m$ and $n$ modes. Let $\bm{r} = ( \bm{r}_A, \bm{r}_B)$ be its first moments and 
\begin{equation}
    V = \begin{bmatrix}
    V_A      & V_{AB} \\ 
    V_{AB}^T & V_B
    \end{bmatrix}
\end{equation}
\noindent be its $2(m+n)\times2(m+n)$ covariance matrix. Then, the reduced density matrix $\rho_A = \mathrm{tr}_B(\rho_{AB})$ describing solely the subsystem $A$ is also a gaussian state with first moments $\bm{r}_A$ and covariance matrix $V_A$.

\subsection*{Logarithmic Negativity}
The choice for employing the logarithmic negativity is due to the fact that it is an entanglement monotone and easily computable for bipartite gaussian states, given their covariance matrix. We denote the LN between the $j$-th and $k$-th modes as $\mathcal{E}_N^{j,k}$, $j,k=\mathrm{cav}, 1, 2, \ldots, N$ and follow the prescription of \cite{Paternostro2007} to evaluate it. First, we extract the $4\times4$ matrix $V^{jk}$ from the total system covariance matrix $V$ by taking its block matrices relative only to the modes $j$ and $k$, which can be written in block form as

\begin{equation}
    V^{j,k} = 
    \begin{bmatrix}
    A & C \\ C^T & B
    \end{bmatrix} \, ,
\end{equation}
\noindent where $A$, $B$, $C$ are $2\times2$ matrices.

Secondly, we calculate the smallest of the symplectic eigenvalues of the matrix $\tilde{V}^{j,k} = (\eye\otimes\sigma_z)V^{jk}(\eye\otimes\sigma_z)$ associated with the partially transposed density matrix for the bipartition with the $j$-th and $k$-th mode \cite{Paternostro2007}. For a bipartite system, this symplectic eigenvalue can be written directly by the form of $V^{j,k}$ given above through \cite{Adesso2007}
\begin{equation}
    \tilde{\nu}_{\rm min} = \sqrt{\sigma/2 - \sqrt{\sigma^2 - 4\det(V)}/2} \, ,
\end{equation}
\noindent where $\sigma = \det(A) + \det(B) - 2\det(C)$, $\eye$ is the $2\times2$ identity matrix and $\sigma_z = \mathrm{diag}(1, -1)$ is a Pauli matrix.

 Finally, the LN is given by:
\begin{equation}
    \mathcal{E}_N^{j,k} =    \mathrm{max}\left[0,-\log( \tilde{\nu}_{\rm min} )\right] \, .
\end{equation}

\subsection*{von Neumann entropy}
The von Neumann entropy $S$ for a gaussian state with associated covariance matrix $V$ is a function of its symplectic eigenvalues $\nu_k$:

\begin{equation}
    S = \sum_{k=1}^{M} g(\nu_k) \quad , \quad g(x) = \frac{x+1}{2}\log(\frac{x+1}{2}) - \frac{x-1}{2}\log(\frac{x-1}{2})
\end{equation}

\noindent where $\nu_k, k=1\ldots M$, can be computed from modulus of the $2M$ eigenvalues of $i \Omega V$ \cite{Lloyd2012}.

\subsection*{Mutual Information}

The mutual information of the total system defined as 
\begin{equation}
\mathcal{I} = \sum_{j=1}^M S_j - S_{\rm total}
\end{equation}

\noindent where $S_{\rm total}$ and $S_j$ respectively denote the von Neumann entropy of the total system and of the $j$-th single mode \cite{Herbut2004}.

\subsection*{Squeezing degree}

In order to quantify the mechanical squeezing generated by the CS interaction, we follow the procedure outlined in \cite{Radim2020}. First we perform a partial trace over the full system in order to arrive at the $2\times2$ covariance matrix $V_j$ describing solely the $j$-th NP; secondly, we quantify the amount of squeezing by finding the variances of the squeezed and antisqueezed quadratures, respectively $V_{\rm sq}$ and $V_{\rm asq}$:

\begin{align}
    V_{\rm sq} = \min( \mathrm{eig}(V_j) ) && V_{\rm asq} = \max( \mathrm{eig}(V_j) ) \, ,
\end{align}
\noindent where $\mathrm{eig}(V_j)$ denotes the eigenvalues of $V_j$. The squeezing degree is, then, $\eta \equiv V_{\rm sq}/V_{\rm asq} \leq 1$.

\subsection*{Numerical simulations of gaussian states}

The numerical simulations for the time evolution and calculations of the theoretical tools for Gaussian states presented above were performed with the custom numerical toolbox in MATLAB developed for this project \cite{numerical_toolbox}.

\section{Decoherence mechanisms} \label{appendix:decoherence}

In this article, we consider two major forms of decoherence/heating for the nanoparticles: thermal decoherence from the collisions with the environmental gas surrounding each NP and recoil heating as each NP incoherently scatters light from its tweezer into free space. 

Current experiments with CS interactions operate in moderate vacuum \cite{Delic2020a, Delic2020b, Windey2019} achieving pressures ranging from $\SI{10^{-6}}{m\bar}$ \cite{Delic2020a} to $\SI{1.4}{m\bar}$ \cite{DeLosRios2020}, at which the residual environmental gas damping on the $j$-th NP is linear in the gas pressure, $p_{\mathrm{gas}, j}$, following \cite{Jain2016, DeLosRios2020},
\begin{equation}
    \gamma_j \approx 15.8 \frac{R_j^2 p_{\mathrm{gas}, j}}{m_j v_{\mathrm{gas}, j} }
\end{equation}
\noindent where $v_{\mathrm{gas}, j} = \sqrt{\frac{3 k_B T\envj}{m_{\rm gas}}}$ is the root-mean-square velocity of a gas molecule with mass $m_{\rm gas}$. 
The associated thermal decoherence rate  from the collisions with the residual gas then becomes $\Gamma_{\mathrm{gas}, j} = \gamma_j \, \overbar{n}_{\mathrm{th}, j}$ \cite{Jain2016}.
Note that lower gas pressures are possible, as demonstrated in \cite{Tebbenjohanns2021, Tebbenjohanns2020}.

In regards to the recoil heating, we will only consider contributions arising from the scattering of photons from the trapping tweezer of each particle. The reasoning for this simplification is twofold. First, as the cavity is not actively driven, its field intensity is much smaller than each tweezer intensity \cite{Delic2020b}, diminishing its contribution. Secondly, as we are considering non-overlapping tweezers, the intensity of the $j$-th OT on the $n$-th NP, $n\neq j$, should be negligible and, thus, disregarded. Therefore, the recoil heating rate on $j$-th NP reads \cite{Jain2016}
\begin{equation}
    \Gamma_{\mathrm{recoil}, j} = \frac{1}{5} \frac{P_{\mathrm{scatt}, j}}{m_j c^2} \frac{\omega\tw}{\omega_j}
\end{equation}
\noindent where $c$ is the speed of light and $P_{\mathrm{scatt}, j} = I_{0, j} \sigma_{\mathrm{scatt}, j}$ is the scattered power of the $j$-th NP from its tweezer; $I_{0,j} = 2P\twj/(\pi w_{0,j}^2)$ is its intensity at its NP's mean position and $\sigma_{\mathrm{scatt}, j} = |\alpha_j|^2 k\tw^4 / (6\pi \epsilon_0^2)$ the scattering cross section of with the NP.

For simplicity, we may consider identical environments surrounding each NP, $p_{\mathrm{gas}, j} = p_{\mathrm{gas}}$ and $T\envj = T_{\rm gas} \, \forall j$, as the residual gas and temperature should a priori be the same everywhere inside the vacuum chamber, resulting in $\Gamma_{\mathrm{gas}, j} = \Gamma_{\mathrm{gas}} \, \forall j$. Furthermore, for the sake of brevity, we can make $\Gamma_{\mathrm{recoil}, j} \approx \Gamma_{\mathrm{recoil}} \, \forall j$ if all the particles are of the same material and size, and the tweezers have approximately the same intensity at their focus.



\section{Coupling strength and coherence time parameter dependence} \label{appendix:experimental_params}


The recent achievement of motional ground state cooling of a NP through Coherent Scattering \cite{Delic2020a} has brought us a step closer to observing entanglement between mesoscopic mechanical objects. As we note in the main text, however, the parameter region accessed by this experiment would not generate such quantum correlations. As a means to this end, we now study how one is able to enhance the CS coupling strength with the constraint of maintaining low decoherence rates and a natural frequency close to the one reported in \cite{Delic2020a}.

When considering our definitions and the density of silica nanospheres, $\rho = \SI{2200}{kg/m^3}$ \cite{Gonzalez-Ballestero2019}, we may write these quantities, omitting the index of the particles for simplicity, as
\begin{align}
    g = \sqrt[4]{\frac{12}{\pi}}\left( \frac{n_{r}^2-1}{n_r^2+2} \right)^{3/4}\, \frac{{P_{t}}^{1/4}\,R^{3/2}\,{\omega _{c}}^{3/2}}{\sqrt{L}\,c^{5/4}\,w_{0,c}} \, \frac{1}{\sqrt[4]{\rho}} && \omega = \sqrt{\frac{12}{\pi}}\left( \frac{n_{r}^2-1}{n_r^2+2} \right)^{1/2} \, \frac{\sqrt{P_t}}{w_0^2} \, \frac{1}{\sqrt{\rho c}}
\end{align}
\vspace{1em}
\begin{align}
    \Gamma_{\mathrm{gas}} = \frac{79\sqrt{3}}{20\pi} \, \frac{\bar{n}_{\mathrm{gas}}\,p_{\mathrm{gas}}}{\rho\sqrt{\frac{k_{B}T_{\mathrm{gas}}}{m_{\mathrm{gas}}}}} \, \frac{1}{R} &&
    \Gamma_{\mathrm{recoil}} = \frac{2\sqrt{3}}{15\sqrt{\pi}}\left( \frac{n_{r}^2-1}{n_r^2+2} \right)^{3/2} \, \frac{\sqrt{P_t}R^3k_t^5}{\sqrt{\rho c}}
\end{align}

\begin{figure}[!ht]
    \centering
    \includegraphics[width=0.75\linewidth]{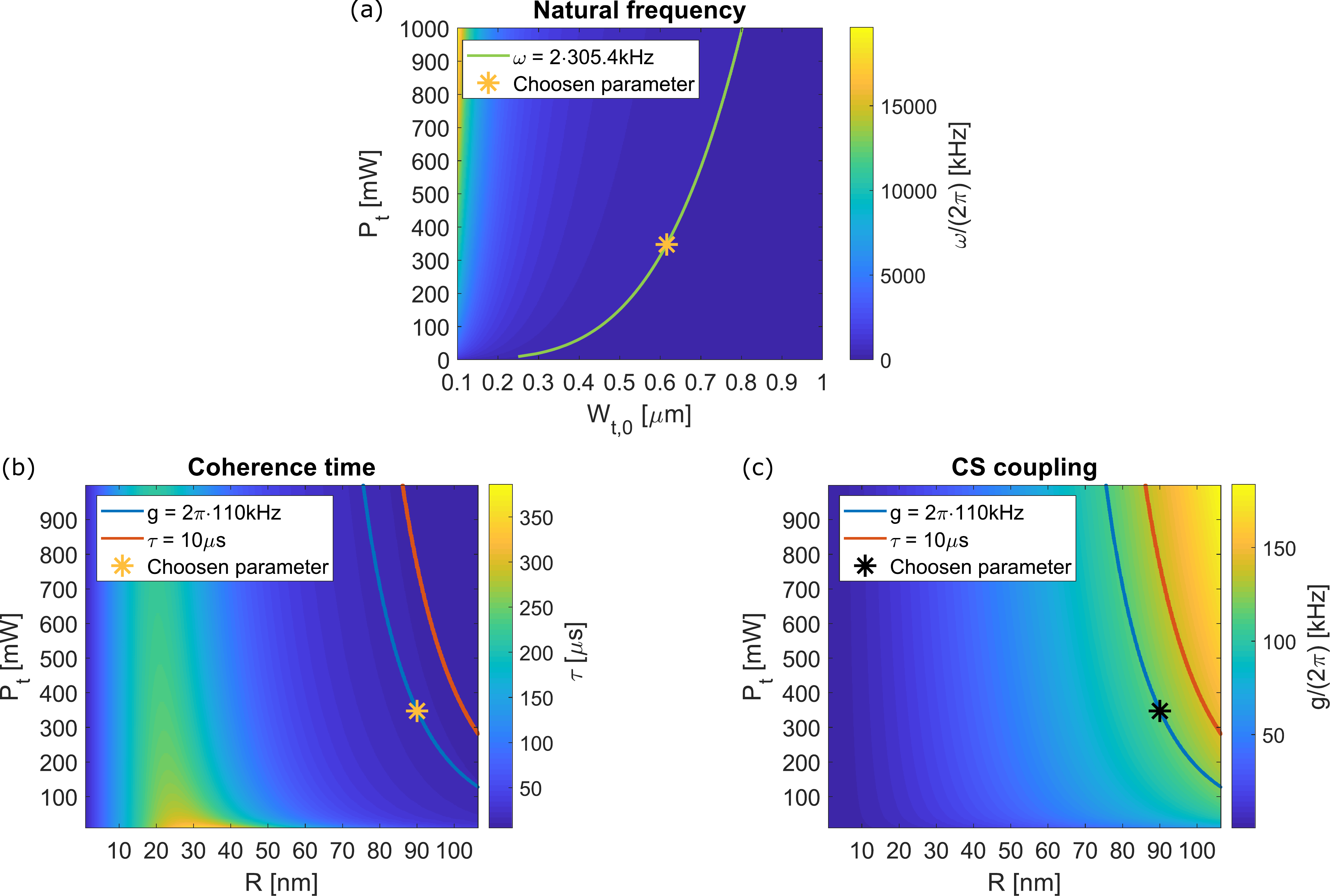}
    \caption{Dependence of (a) mechanical frequency, (b) coherence time and (c) coupling strength in respect to the tweezer's power and waist, and particle radius. Every other parameter is given in Table \ref{tab:experimental_values_delic}. Note the region comprised between the light blue curve ($g= 2\pi \cdot \SI{110}{kHz}$) and the red curve ($\tau = \SI{10}{\mu s}$) exhibits both high coupling and coherence time when compared to current experimental realizations.}
    \label{fig:freq_coupling_coherence_time}
\end{figure}

We present these full expressions in order to show that while the coherence time and coupling strength have a dependence with the tweezer power, waist and particle radius, the frequency does not depend on the latter. If we are to set a target frequency of $\omega = \SI{2\pi\cdot305.4}{kHz}$, we may relate the tweezer power to its waist, and thus only require to study the dependence of these variables upon $R$ and $P_t$. In Figure \ref{fig:freq_coupling_coherence_time}(a), we use the parameters given in Table \ref{tab:experimental_values_delic} to study how the mechanical frequency depends on the particle size and tweezer power. In parts (b) and (c) of this figure, using the same parameters, we show how the coupling strength $g$ and the coherence time $\tau = 1/(\Gamma_{\mathrm{gas}} + \Gamma_{\mathrm{recoil}})$ vary as a function of $R$ and $P_{\mathrm{t}}$, alongside their cross sections at $g= 2\pi \cdot \SI{110}{kHz}$ and $\tau = \SI{10}{\mu s}$, respectively in light blue and red. 
Thus, between these two curves, we have a parameter region with sufficiently high coupling strength and enough coherence time such that mechanical entanglement should be able to occur. Taking into account Figure \ref{fig:log_neg_coupling_time}, experimental realizations on this region should be able to give birth to mechanical entanglement if the nanoparticles are sufficiently pre-cooled down. 

We choose the parameters in the point marked by the asterisk as it defines a stable system with a steady state that also allows for numerical stability of our simulations. The chosen values are: $R = \SI{90}{nm}$, $P_t = \SI{347.5}{mW}$, $w_{0} = \SI{0.61583}{\mu m}$, $T_{\mathrm{gas}} = \SI{300}{K}$. This results in the simulation parameters presented in Table \ref{tab:proposed_parameters} with coherence time $\tau = \SI{14.816}{\mu s}$. For comparison below, Ref. \cite{Delic2020a} used the parameters given in Table \ref{tab:experimental_values_delic} and observed $\Gamma_{\mathrm{gas},0} = 2\pi\times \SI{16.1}{\kHz}$ and $\Gamma_{\mathrm{recoil},0} = 2\pi\times \SI{6}{\kHz}$ resulting in a maximum coherence time of $\tau_0 = 1/(\Gamma_{\mathrm{gas},0} + \Gamma_{\mathrm{recoil},0}) \approx \SI{7.6}{\mu s} $ corresponding to approximately $15$ oscillations. Thus, we have found realistic parameters which allow for both higher coupling strength and coherence time.



\end{document}